\documentclass[pra,reprint,a4paper,showpacs]{revtex4-1}
\usepackage{bm}
\usepackage{ulem}
\usepackage[dvipdfmx]{graphicx}
\usepackage{dcolumn}
\usepackage{color}
\usepackage{amsmath}

\def\labelenumi{\theenumi}

\draft 

\begin{document}


\title{Time-dependent multiconfiguration self-consistent-field method
based on occupation restricted multiple active space model
for multielectron dynamics in intense laser fields
}



\author{Takeshi Sato}
\email[Electronic mail:]{sato@atto.t.u-tokyo.ac.jp}
\author{Kenichi L. Ishikawa}
\email[Electronic mail:]{ishiken@atto.t.u-tokyo.ac.jp}
\affiliation{
Photon Science Center, School of Engineering, 
The University of Tokyo, 7-3-1 Hongo, Bunkyo-ku, Tokyo 113-8656, Japan
}
\affiliation{
Department of Nuclear Engineering and Management, School of Engineering,
The University of Tokyo, 7-3-1 Hongo, Bunkyo-ku, Tokyo 113-8656, Japan
}



\begin{abstract}
The time-dependent multiconfiguration self-consistent-field method based
on the occupation-restricted multiple active space model is
proposed (TD-ORMAS) for multielectron dynamics in intense laser fields.
Extending the previously proposed
time-dependent
complete-active-space self-consistent-field  method [TD-CASSCF; Phys.~Rev.~A, {\bf 88}, 023402 (2013)], which
divides the occupied orbitals into core and active orbitals, the
TD-ORMAS method {\it further} subdivides the active orbitals into an
arbitrary number of subgroups, and poses the {\it occupation restriction} by
giving the minimum and maximum number of electrons distributed in each
subgroup. This enables highly flexible construction of the
configuration interaction (CI) space, 
allowing a large-active-space simulation of dynamics, e.g.,
the core excitation or ionization.
The equations of motion both for CI coefficients and spatial
orbitals are derived based on the time-dependent variational principle,
and an efficient algorithm is proposed to solve for the orbital time derivatives.
In-depth descriptions of the computational implementation are given in a
readily programmable manner.
The numerical application to the one-dimensional lithium hydride cluster models
demonstrates that the high flexibility of the TD-ORMAS framework allows for
the cost-effective simulations of multielectron dynamics, by exploiting
systematic series of approximations to the TD-CASSCF method.
\end{abstract}


\maketitle 

\section{introduction\label{sec:introduction}}
One of the main objectives of strong field physics and attosecond
science is a direct measurement and control of
electron motions in atoms and molecules \cite{Krausz2009RMP}. 
The time-dependent Schr\"odinger equation (TDSE) provides the rigorous
theoretical framework for investigating such electron dynamics 
\cite{Pindzola1998PRA,Pindzola1998JPB,Colgan2001JPB,
Parker2001,Laulan2003PRA,Piraux2003EPJD,Laulan2004PRA,ATDI2005,
Feist2009PRL,Pazourek2011PRA,He_TPI2012PRL,Suren2012PRA,He_TPI2013AS,
Vanroose2006PRA,Horner2008PRL,Lee2010JPB}.
However, direct real-space simulations
of \textcolor{black}{the} TDSE to systems with more than two electrons are extremely
difficult. To investigate  multielectron dynamics in intense laser fields, 
the multiconfiguration time-dependent Hartree-Fock (MCTDHF) method has
been developed \cite{Caillat2005PRA,Kato:2004,Nest:2005a,Haxton:2011,Hochstuhl:2011}, in which the
time-dependent total 
wavefunction is given in the configuration interaction (CI) expansion,
\begin{eqnarray}\label{eq:mcscf}
 \Psi(t) = \sum_{\bf I} \Phi_{\bf I}(t) C_{\bf I}(t),
\end{eqnarray}
where $\Phi_{\bf I}(t)$ is a Slater determinant built from a given
number, $n$, of orbital functions $\{\phi_i(t)\}$.
Both CI coefficients $\{C_{\bf I}\}$ and orbitals are
simultaneously varied in time, which allows 
the use of considerably smaller number of orbitals than in the fixed
orbital approach.
This method, however, suffers from the exponential increase of the
computational cost against the number of electrons $N$.

To circumvent this difficulty, we have recently proposed the time-dependent
complete-active-space self-consistent-field (TD-CASSCF) method
\cite{Sato:2013}, which divides the orbitals into core and active
orbitals. 
Simultaneously, the total electrons are classified into core
and active electrons, $N = N_\textrm{C} + N_\textrm{A}$, and the CI
expansion of Eq.~(\ref{eq:mcscf}) 
consists of all Slater determinants
including doubly occupied core orbitals. 
The flexible core-active classification enables compact yet accurate
representation of dynamics according to the given physical situation.
For example, in the presence of intense, long-wavelength laser pulse\textcolor{black}{s}, the tightly bound electrons are
expected to remain nonionized, while only weekly bound electrons ionize
appreciably. The TD-CASSCF method is ideally suited to such situations,
with tightly- and weakly-bound electrons treated as core and active, 
respectively. 

The TD-CASSCF method is featured by the {\it fully} correlated
description of the active electrons, by means of the {\it complete} CI
expansion within the active orbitals. This guarantees that important
properties of the rigorous MCTDHF method are preserved for the TD-CASSCF
method including core orbitals \cite{Sato:2013}.
However, the complete CI expansion still results in the
exponential scaling of the computational cost albeit with respect to
$N_\textrm{A}$, not to $N$.
This causes an immediate difficulty when, e.g., the core
ionization from tightly bound orbitals is of interest. 
In such situations, \textcolor{black}{all} electrons would have to be assigned as
active.
One should also recognize that, even for
the dynamics dominated by chosen active electrons, the TD-CASSCF result with
small number, $n_\textrm{A}$, of active orbitals is at best qualitative.
Instead \textcolor{black}{a} sufficiently large number of orbitals,
typically $n_\textrm{A} \geq 2N_\textrm{A}$ \textcolor{black}{is}
required to obtain quantitatively, or even qualitatively correct
descriptions. Again, the computational cost grows steeply against
$n_\textrm{A}$ for a fixed $N_\textrm{A}$, hindered large-active-space
calculations. Clearly, the {\it non-complete} CI expansion is mandatory
to have wider range of problems at hand. 

An important step in this direction has been made in
Ref.~\cite{Miyagi:2013, Miyagi:2014b}, 
which divides the active orbitals into two subspaces, and allows
variable distributions of electrons among the two subspaces. 
The method was applied to the one-dimensional model Hamiltonian \cite{Miyagi:2014b}, with
the total wavefunction given 
by the truncated CI expansion;
\begin{eqnarray}\label{eq:rhf+_1stq}
\Psi  = \Phi_0 C_0 + \sum_{ia}\Phi_i^a C_i^a +
\sum_{ijab} \Phi_{ij}^{ab} C_{ij}^{ab} + \cdot\cdot\cdot, 
\end{eqnarray}
(time argument is omitted) where $\Phi_0$ is the closed-shell
Hartree-Fock determinant built from the $N_\textrm{A}/2$ spatial
orbitals, $\Phi_i^a$ is the singly excited determinant with $\phi_i$ in
$\Phi_0$ replaced by $\phi_a$ in the second active subspace,
$\Phi_{ij}^{ab}$ is the analogous doubly excited determinant, etc, 
truncated after a given order of excitations.
Although 
this wavefunction converges to the
complete-CI wavefunction with up to $N_\textrm{A}$-fold excitations
included, the accuracy of this method is strongly system dependent, 
as 
discussed
in the present work. In addition, the computational 
algorithm proposed in Ref.~\cite{Miyagi:2014b} involves severe
bottleneck in increasing the order of excitations in
Eq.~(\ref{eq:rhf+_1stq}). More flexible and efficient method is required
to take full advantage of, and minimize the drawback of non-complete CI
expansions.

In this work, 
we adopt the occupation-restricted multiple-active-space 
(ORMAS) model \cite{Ivanic:2003a}, originally developed for stationary
electronic structure problems, as a highly flexible framework to construct
non-complete CI spaces.
On top of the core-active subspacing, the ORMAS method
{\it further} divides the active orbitals into an arbitrary 
number of subgroups, and poses the {\it occupation restriction} through specifying
the minimum and maximum numbers of electrons distributed in each subgroup.
%
The ORMAS method has been applied \cite{Ivanic:2003a, Ivanic:2003b}
both to fixed-orbital CI methods and to the multiconfiguration
self-consistent-field (MCSCF) method, where not 
only CI coefficients but also the occupied orbitals are variationally 
optimized. Our interest is placed on the latter, in the context of the
time-dependent non-stationary problems. Namely, we develop the
time-dependent MCSCF method based on the ORMAS
model, hereafter called the TD-ORMAS method.

This paper proceeds as follows. In Sec.~\ref{sec:ansatz}
the ORMAS method is introduced in the rigorous second quantization
formalism. Then in Sec.~\ref{sec:td-ormas}, the equation of motion for
the TD-ORMAS method is derived based on the time-dependent variational
principle. The computational implementation is described in detail in
Sec.~\ref{sec:implementation}. 
The performance of the TD-ORMAS method is assessed using
one-dimensional multielectron models
in Sec.~\ref{sec:applications}. Finally, concluding
remarks are given in Sec.~\ref{sec:summary}. Appendix~\ref{app:amat_bvec}--\ref{app:imag},
and \ref{app:ormas}, respectively, describe further details of theory and
implementation, and another numerical example.
The Hartree atomic units are used throughout unless otherwise noted.

\section{Ansatz\label{sec:ansatz}}
In this section, we introduce the ORMAS method \cite{Ivanic:2003a}.
Since we consistently rely on the second quantization formalism in this work,
we first briefly discuss the second quantized representation of the
MCSCF wavefunction, followed by the rigorous definition of the ORMAS
method. We consider a system with $N_\uparrow$ ($N_\downarrow$) up
(down) spin electrons, thus $N = N_\uparrow + N_\downarrow$ total electrons.

\subsection{MCSCF wavefunctions in the second quantization\label{subsec:mcscf}}
We define the set of $N_b$ orthonormal spatial orbitals,
$\{\phi_\mu\}$, assumed to span the spinless one-electron Hilbert space $\mathcal{H}$.
In principle, $\mathcal{H}$ consists of infinite number of orbitals, but
in practice, the number of orbitals $N_b$ is determined by the number of
underlying basis functions, e.g., the number of spatial grid points in the finite
difference approach. 
The one-electron complete-orthonormal basis is constructed by the direct product
$\mathcal{H}\times\{\uparrow, \downarrow\}$, where $\uparrow$
($\downarrow$) represents the up (down) spin eigenfunction. This implies
the spin-restricted treatment, using the same spatial orbitals for up
and down spin orbitals. For each element of
$\mathcal{H}\times\{\uparrow, \downarrow\}$, the Fermion creation 
(annihilation)
operator $\hat{a}^\dagger_{\mu\sigma}$ ($\hat{a}_{\mu\sigma}$) is
associated, with $\sigma \in \{\uparrow,\downarrow\}$.

The MCSCF wavefunction is based on the division of the full Hilbert space
$\mathcal{H}$ into occupied ($\mathcal{P}$) and virtual ($\mathcal{Q}$)
orbital subspaces,
\begin{eqnarray}
 \mathcal{H} &=& \mathcal{P} + \mathcal{Q},
\end{eqnarray}
where $\mathcal{P}$ has $n$ members called {\it occupied} orbitals,
and remaining {\it virtual} orbitals form the $\mathcal{Q}$ space:
\begin{eqnarray}
\mathcal{P} &=& \left\{\phi_1,\phi_2, \cdot\cdot\cdot, \phi_n\right\}, \\
\mathcal{Q} &=& \left\{\phi_{n+1}, \phi_{n+2}, \cdot\cdot\cdot \right\}.
\end{eqnarray}
The determinant $\Phi_{\bf I}$ of Eq.~(\ref{eq:mcscf}) is built
from the $\mathcal{P}$ space orbitals only. The essence of the
MCSCF method, both in the time-dependent and time-independent
theories, is the variational separation of $\mathcal{P}$ and
$\mathcal{Q}$ spaces; the CI problem is solved within the optimized
$\mathcal{P}$ space.

It is possible, and highly beneficial \cite{Sato:2013}, to separate the
occupied space into core ($\mathcal{C}$) and active ($\mathcal{A}$) subspaces,
\begin{eqnarray}
\mathcal{P} &=& \mathcal{C} + \mathcal{A},
\end{eqnarray}
where $\mathcal{C}$ consists of $n_\textrm{C}$ {\it core} orbitals, and
$\mathcal{A}$ of $n_\textrm{A}$ {\it active} orbitals, with $n =
n_\textrm{C} + n_\textrm{A}$:
\begin{eqnarray}
\mathcal{C} &=& \left\{\phi_1,\phi_2, \cdot\cdot\cdot, \phi_{n_\textrm{C}}\right\}, \\
\mathcal{A} &=& \left\{\phi_{n_\textrm{C}+1}, \phi_{n_\textrm{C}+2}, \cdot\cdot\cdot, \phi_n\right\}.
\end{eqnarray}
At the same time, $N$ electrons are classified into $N_\textrm{C}$ core
electrons and $N_\textrm{A}$ active electrons, where
\begin{eqnarray}
N_\textrm{C} &=& 2 n_\textrm{C}, \\
N_\textrm{A} &=& N - N_\textrm{C}.
\end{eqnarray}
With these relations, the summation ${\bf I}$ 
in Eq.~(\ref{eq:mcscf}) is taken over those Slater determinants
including $n_\textrm{C}$ doubly occupied core orbitals. 
Thus in the second quantization we write
\begin{eqnarray}\label{eq:mcscf_2q}
|\Psi\rangle = \hat{\Phi}_\textrm{C} |\Psi_\textrm{A}\rangle, \hspace{.5em}
|\Psi_\textrm{A}\rangle = \sum_{\bf I}^{\sf P} |{\bf I}\rangle C_{\bf I},
\end{eqnarray}
where $\hat{\Phi}_\textrm{C} \equiv \prod_{i\in\mathcal{C}}
\hat{a}^\dagger_{i\uparrow} \hat{a}^\dagger_{i\downarrow}$ and
$|{\bf I}\rangle$ represent the core and active parts of the 
determinant $\Phi_{\bf I}$ in Eq.~(\ref{eq:mcscf}), respectively,
with
\begin{eqnarray}
\label{eq:det_act}
\vert {\bf I} \rangle &=& \hat{\bf I}\vert\rangle,
\hspace{.5em}
\hat{\bf I} = \prod_\sigma\prod_{t \in \mathcal{A}}
(\hat{a}^\dagger_{t\sigma})^{I_{t\sigma}},
\end{eqnarray}
where $|\rangle$ represents the vacuum state,
$I_{t\sigma} = \{0,1\}$, and $\sum_\sigma\sum_{t \in \mathcal{A}}
I_{t\sigma} = N_\textrm{A}$.
In Eq.~(\ref{eq:mcscf_2q}), the summation {\bf I} runs through the
element of a CI space ${\sf P}$, which in general consists of
a given set of {\color{black}active determinants $\{|{\bf I}\rangle\}$}. 
Up to now, Eq.~(\ref{eq:mcscf_2q}) represents the general MCSCF
wavefunction ($n_\textrm{C}$ can be zero). 
We separate the core part in Eq.~(\ref{eq:mcscf_2q})
to maximally exploit the simplification due to the core wavefunction.
{\color{black}
In what follows, the term determinant denotes the active part $|{\bf I}\rangle$.
}

For later convenience, we introduce the following symbols:
\begin{eqnarray}
\hat{\bf I}(\mathcal{A}^\prime) = \prod_\sigma\prod_{t \in \mathcal{A}^\prime}
(\hat{a}^\dagger_{t\sigma})^{I_{t\sigma}},
\end{eqnarray}
\begin{eqnarray}
\left[\mathcal{A}^\prime\right]^{N^\prime} \equiv
\left\{\hat{\bf I}(\mathcal{A}^\prime);
\sum_\sigma\sum_{t \in \mathcal{A}^\prime}I_{t\sigma} = N^\prime\right\},
\end{eqnarray}
where $\hat{\bf I}(\mathcal{A}^\prime)$ is the segment of $\hat{\bf
I}$ for a given subset of active orbitals
$\mathcal{A}^\prime \subset \mathcal{A}$, and
$[\mathcal{A}^\prime]^{N^\prime}$ denotes
the set of determinants constructed by distributing a given number,
$N^\prime$, of electrons among orbitals in $\mathcal{A}^\prime$ in all
the possible ways. 

\subsection{ORMAS wavefunction\label{subsec:ormas}}
In the ORMAS model \cite{Ivanic:2003a},
the active orbital space $\mathcal{A}$ is {\it further} subdivided into a
given number, $G$, of subgroups;
\begin{eqnarray}
\mathcal{A} &=& \mathcal{A}_1 + \mathcal{A}_2 + \cdot\cdot\cdot + \mathcal{A}_G,
\end{eqnarray}
\begin{eqnarray}
\mathcal{A}_1 &=&
\left\{\phi^{(1)}_1,\phi^{(1)}_2,\cdot\cdot\cdot,\phi^{(1)}_{n_1}\right\},
\nonumber \\
\mathcal{A}_2 &=&
\left\{\phi^{(2)}_1,\phi^{(2)}_2,\cdot\cdot\cdot,\phi^{(2)}_{n_2}\right\},
\nonumber \\
&\cdot\cdot\cdot& \nonumber \\
\mathcal{A}_G &=&
\left\{
\phi^{(G)}_1\hspace{-.25em},
\phi^{(G)}_2\hspace{-.25em},\cdot\cdot\cdot,
\phi^{(G)}_{n_G}\hspace{-.25em}\right\},
\end{eqnarray}
with $n_\textrm{A} = \sum_{g=1}^G n_g$,
and $\phi^{(g)}_{j} \equiv \phi_i; i = n_\textrm{C} + \sum_{g^\prime = 1}^{g-1}n_{g^\prime}+j$.
At the same time, the {\it occupation restriction} is posed through
specifying the minimum and maximum numbers of electrons in each subgroup; 
\begin{eqnarray}\label{eq:occb}
N^\textrm{min}_1 \leq &N_1& \leq N^\textrm{max}_1, \nonumber \\
N^\textrm{min}_2 \leq &N_2& \leq N^\textrm{max}_2, \nonumber \\
&\cdot\cdot\cdot& \nonumber \\
N^\textrm{min}_G \leq &N_G& \leq N^\textrm{max}_G,
\end{eqnarray}
with
\begin{eqnarray}\label{eq:occsum}
N_\textrm{A} = N_1 + N_2 + \cdot\cdot\cdot + N_G.
\end{eqnarray}
The boundaries of Eq.~(\ref{eq:occb})
determine the possible set of {\it occupation distributions}
$\bm{d} = (N_1,N_2,\cdot\cdot\cdot,N_G)$ which satisfies Eq.~(\ref{eq:occsum}).
Upon this active subspacing and occupation restriction, the ORMAS-CI
space is constructed as
\begin{eqnarray}\label{eq:cip_mas}
{\sf P}_\textrm{ORMAS} = \sum_{\bm{d}} {\sf P}(\bm{d}),
\end{eqnarray}
\begin{eqnarray}\label{eq:cip_dist} 
&&{\sf P}(\bm{d}) =
\label{eq:cip_dist1}
\left[\mathcal{A}_1\right]^{N_1}
\left[\mathcal{A}_2\right]^{N_2} \cdot\cdot\cdot
\left[\mathcal{A}_G\right]^{N_G} \\ &&=
\label{eq:cip_dist2} 
\left\{\hat{\bf I}=\hat{\bf I}_1\hat{\bf I}_2\cdot\cdot\cdot\hat{\bf I}_G; \hspace{.5em}
\sum_\sigma\sum_{t \in \mathcal{A}_g} I_{t\sigma} = N_g, 1 \leq g \leq G
\right\}, \nonumber \\
\end{eqnarray}
where $\hat{\bf I}_g \equiv \hat{\bf I}(\mathcal{A}_g)$.
The ORMAS-CI space is given by the direct sum [Eq.~(\ref{eq:cip_mas})]
of disjoint CI spaces ${\sf P}(\bm{d})$ for all the allowed distributions
$\{\bm{d} = (N_1,N_2,\cdot\cdot\cdot,N_G)\}$, where ${\sf P}(\bm{d})$ 
is the direct product space [Eq.~(\ref{eq:cip_dist1})], consisting of determinants
built by distributing $N_1$ electrons in the subgroup $\mathcal{A}_1$,
$N_2$ electrons in $\mathcal{A}_2$, $\cdot\cdot\cdot$, and $N_G$
electrons in $\mathcal{A}_G$, in all the possible ways [Eq.~(\ref{eq:cip_dist2})].

\begin{figure}[!t]
\centering
\includegraphics[width=.90\linewidth,clip]{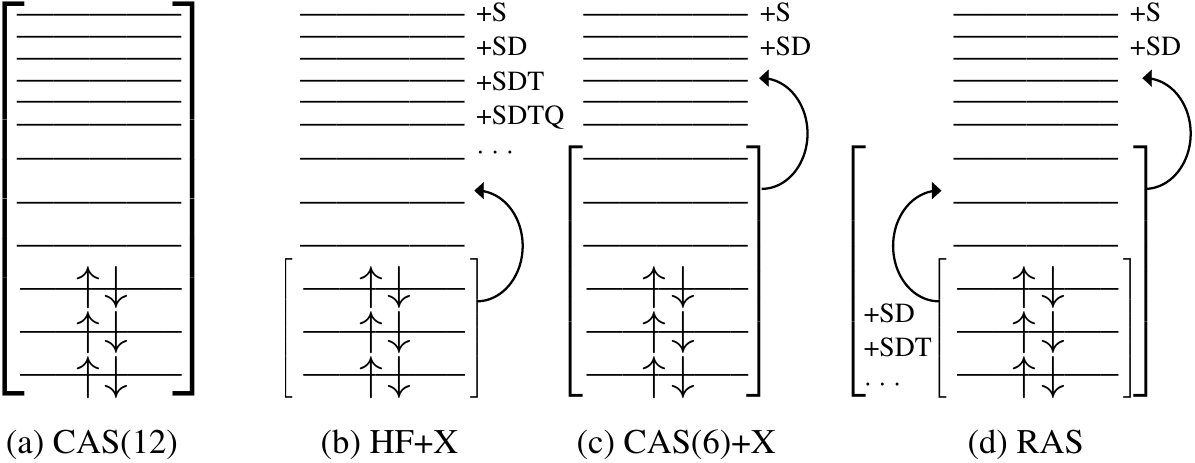}
\caption{\label{fig:ormas}Examples of the ORMAS-CI space for
6 active electrons and 12 active orbitals. 
(a) The CAS(12) space with no subdivision of active space.
(b) The Hartree-Fock reference CI space with $(n_1,n_2)=(3,9)$.
(b) The CAS(6) reference CI space with $(n_1,n_2)=(6,6)$. 
(c) An example of the RAS CI space with $(n_1,n_2,n_3)=(3,3,6)$.
The first (lowest) group of orbitals, and also the union of
 first two groups in 
the case of RAS CI space, are bracketed. The straight up and down
arrows represent electrons in the HF configuration, which are to be distributed
according to the respective ORMAS restriction. The curved upward arrows
image the excitations from one to the other subgroup. See text for more
details.
}%
\end{figure}
Here we give a few examples of the ORMAS-CI spaces:
\begin{enumerate}
\renewcommand{\labelenumi}{\arabic{enumi})}
\item No subdivision of $\mathcal{A}$
      ($G=1, N^1_\textrm{min}=N^1_\textrm{max}=N_\textrm{A}$)
      gives the {\color{black}complete CI space 
      } [Fig.~\ref{fig:ormas}~(a)];
      \begin{eqnarray}\label{eq:cip_cas}
       {\sf P}_\textrm{CAS} = \left[\mathcal{A}\right]^{N_\textrm{A}}.
      \end{eqnarray}
      The TD-CASSCF method \cite{Sato:2013} is based on this CI space. 
      In this work, the TD-CASSCF method with $n_\textrm{A}$ active
      orbitals is simply denoted as CAS($n_\textrm{A}$). 
\item As shown in Fig.~\ref{fig:ormas}~(b) and (c), dividing $\mathcal{A}$ into
      two subgroups ($G=2$), and restricting the occupation by 
      $N_\textrm{A}-L \leq N_1 \leq N_\textrm{A}$, $0 \leq N_2 \leq L$, 
      for a given $L$ generates
      the CI space including all the determinants built from
      the first (lowest in Fig.~\ref{fig:ormas}) active
      orbitals ({\it reference CI space}), and those 
      configurations generated by single, double, triples,
      $\cdot\cdot\cdot$, and up to $L$-fold excitations from the
      reference to the second subgroup. Especially, we focus on the
      following two schemes;

      \noindent{\it Hartree-Fock reference CI space}:
      With $n_1 = N_\textrm{A}/2$, the CI space includes the
      Hartree-Fock reference determinant plus excitations from the
      reference to the second subgroup. The corresponding TD-ORMAS
      method is denoted as HF+X for brevity, with X = S, SD, SDT, and so on, 
      indicating the inclusion of only single, single and double, 
      up to triple excitations, etc. The first quantized expression is
      given by Eq.~(\ref{eq:rhf+_1stq}), which is pictorially explained
      in Fig.~\ref{fig:ormas}~(b)

      \noindent{\it CAS($N_\textrm{A}$) reference CI space}:
      With $n_1 = N_\textrm{A}$, the CI space consists of
      all the determinants built from the first $N_\textrm{A}$ active
      orbitals [CAS($N_\textrm{A})$ reference], plus
      excitations from the reference to the second subgroup. The
      corresponding TD-ORMAS method is denoted as CAS($N_\textrm{A}$)+X. 
      An example is given in Fig.~\ref{fig:ormas}~(c)
\item Another important example, shown in Fig.~\ref{fig:ormas}~(d), 
      is the so-called {\it restricted active space} (RAS) model proposed 
      by Olsen {\it et al} \cite{Olsen:1988}, which divides $\mathcal{A}$ into
      three subgroups ($G=3$), and restricts the occupation by setting
      the maximum number of holes $M_\text{hole}$ in $\mathcal{A}_1$ and
      the maximum number of electrons $M_\text{elec}$ in
      $\mathcal{A}_3$, while $N_2$ is unconstrained. 
      In the ORMAS notation, this corresponds to the following boundaries;
      \begin{eqnarray}\label{eq:or_ras}
       N_\textrm{A} - M_\textrm{hole} \leq N_1 \leq N_\textrm{A}, 
	\hspace{.5em} 0 \leq N_3 \leq M_\textrm{elec}.
      \end{eqnarray}
      The RAS scheme allows excitations (1) from the first subgroup into the second,
      and (2) from the union of first two subgroups into the third, up to
      different maximum ranks, $M_\textrm{hole}$ and $M_\textrm{elec}$
      for (1) and (2), respectively.
      {\color{black} Figure~\ref{fig:ormas}~(d) shows a special case with
      $N_\textrm{A} = 2n_1$, 
      for which $M_\textrm{elec} \leq M_\textrm{hole}$ should hold.}

\end{enumerate}
Note that the ``TD-RASSCF'' method proposed in Ref.~\cite{Miyagi:2014b}
uses the second type of CI spaces, but not based on the RAS scheme
of Ref.~\cite{Olsen:1988}. To avoid confusion
and for consistency with the terminology widely used in the
stationary electronic structure theory,
we refer to the latter method as RAS,
which includes the method of Ref.~\cite{Miyagi:2014b} as a
special case.
The ORMAS framework can be used to construct
a variety of other CI spaces as summarized in Ref.~\cite{Ivanic:2003a},
allowing a tailored approximation for a given problem.
In Sec.~\ref{sec:applications} we discuss the physical significance and
computational (dis)advantages of these models with numerical applications.

\section{TD-ORMAS method\label{sec:td-ormas}}
\subsection{Time-dependent variational method in the second quantization\label{subsec:tdvp}}
We first review
the general EOMs for CI coefficients and orbitals [Eq.~(\ref{eq:g-tdci})
and (\ref{eq:g-tdmo}) below] derived in our previous work \cite{Sato:2013}. 
The same equations have been the basis of Refs.~\cite{Miranda:2011a,
Miyagi:2013, Miyagi:2014b}. Based on the time-dependent variational
principle \cite{Frenkel:1934, 
Lowdin:1972, Moccia:1973}, the following action integral $S$,
\begin{eqnarray}\label{eq:s}
S = \int\!dt \langle\Psi| \hat{H} - {\rm  i}\frac{\partial}{\partial t} |\Psi\rangle,
\end{eqnarray}
is required to be stationary, i.e., $\delta S = 0$, with
\begin{eqnarray}\label{eq:ds}
\delta S = \delta \langle\Psi|\hat{H}|\Psi\rangle - {\rm  i}\left(
\langle\delta\Psi|\frac{\partial\Psi}{\partial t}\rangle -
\langle\frac{\partial\Psi}{\partial t}|\delta\Psi\rangle
\right).
\end{eqnarray}
Here $\hat{H}$ is the spin-free second-quantized Hamiltonian,
\begin{eqnarray}
\label{eq:ham}
\hat{H} = \sum_{\mu\nu} h^\mu_\nu \hat{E}^\mu_\nu
+ \frac{1}{2} \sum_{\mu\nu\lambda\gamma} g^{\mu\lambda}_{\nu\gamma} \hat{E}^{\mu\lambda}_{\nu\gamma},
\end{eqnarray}
with
$
\hat{E}^\mu_\nu = \sum_\sigma
\hat{a}^\dagger_{\mu\sigma}\hat{a}_{\nu\sigma}
$, 
$
\hat{E}^{\mu\lambda}_{\nu\gamma} = \sum_{\sigma\tau}
\hat{a}^\dagger_{\mu\sigma}\hat{a}^\dagger_{\lambda\tau}
\hat{a}_{\gamma\tau}\hat{a}_{\nu\sigma}
$, and
\begin{eqnarray}
\label{eq:ham1e}
h^\mu_\nu = \int d\bm{r} \phi^*_\mu(\bm{r}) 
h\left(\bm{r}, \bm{\nabla}_r\right)
\phi_\nu(\bm{r}),
\end{eqnarray}
\begin{eqnarray}
\label{eq:ham2e}
g^{\mu\lambda}_{\nu\gamma} = \int\!\!\!\int \! d\bm{r}_1 d\bm{r}_2
\frac{
\phi^*_\mu(\bm{r}_1) \phi_\nu(\bm{r}_1)
\phi^*_\lambda(\bm{r}_2) \phi_\gamma(\bm{r}_2)
}{|\bm{r}_1 - \bm{r}_2|},
\end{eqnarray}
where the one-electron matrix element $h^\mu_\nu$ consists of kinetic,
nucleus-electron, and external laser terms.
The orthonormality-conserving representation
of variations and time derivatives of orbitals are given
\cite{Miranda:2011a, Sato:2013} by
\begin{eqnarray}
\label{eq:varmo}
\delta\phi_p &=& \sum_\mu \phi_\mu \Delta^\mu_p, \hspace{.5em}
\Delta^\mu_p = \langle\phi_\mu|\delta\phi_p\rangle,
\\
\label{eq:dermo}
\frac{\partial\phi_p}{\partial t} &=& \sum_\mu \phi_\mu X^\mu_p, \hspace{.5em}
X^\mu_p = \langle\phi_\mu|\frac{\partial\phi_p}{\partial t}\rangle,
\end{eqnarray}
in terms of anti-Hermitian transformation matrices $\Delta$ and $X$.
Note that in Ref.~\cite{Sato:2013}, the Hermitian matrix $R \equiv {\rm
i}X$ was used as the working variable. We change notation for a better
transferability between real and imaginary time equations as discussed
in appendix~\ref{app:imag}.
Using these matrices, the variation $\delta\Psi$ and the time derivative
$\dot{\Psi}\equiv \partial\Psi/\partial t$ of the total wavefunction are
compactly given \cite{Sato:2013} by; 
\begin{subequations}\label{eq:varder}
\begin{eqnarray}
\label{eq:var}
|\delta\Psi\rangle &=& 
\hat{\Phi}_\textrm{\color{black}C}\textcolor{black}{\sum_{\bf I}}
|{\bf I}\rangle\delta C_{\bf I} + \hat{\Delta}|\Psi\rangle \\
\label{eq:der}
|\dot\Psi\rangle &=& 
\hat{\Phi}_\textrm{\color{black}C}\textcolor{black}{\sum_{\bf I}}
|{\bf I}\rangle\dot{C}_{\bf I} + \hat{X}|\Psi\rangle,
\end{eqnarray}
\end{subequations}
where $\hat{\Delta}=\sum_{\mu\nu}{\color{black}\Delta}^\mu_\nu\hat{E}^\mu_\nu$, 
$\hat{X}=\sum_{\mu\nu}X^\mu_\nu\hat{E}^\mu_\nu$.
Inserting Eqs.~(\ref{eq:varder}) into Eq.~(\ref{eq:ds}) and requiring 
$\delta S/\delta C^*_{\bf I} = 0, \delta S/\delta \Delta^\mu_\nu = 0$,
gives \cite{Sato:2013}
\begin{eqnarray}
\label{eq:g-tdci}
\dot{C}_{\bf I} = - {\rm i}\langle{\bf I}| 
\hat{\Phi}^\dagger_\textrm{C}
\hat{H} |\Psi\rangle
-
\langle{\bf I}| 
\hat{\Phi}^\dagger_\textrm{C}
\hat{X} |\Psi\rangle,
\end{eqnarray}
\begin{eqnarray}
\label{eq:g-tdmo}
\langle\Psi|
\hat{E}^\mu_\nu\hat{{\sf Q}}\hat{X} -
\hat{X}\hat{{\sf Q}}\hat{E}^\mu_\nu
|\Psi\rangle = -{\rm i}
\langle\Psi|
\hat{E}^\mu_\nu\hat{{\sf Q}}\hat{H} -
\hat{H}\hat{{\sf Q}}\hat{E}^\mu_\nu
|\Psi\rangle. \nonumber \\
\end{eqnarray}
Hereafter, we use notations
$\hat{{\sf P}}$ and $\hat{{\sf Q}}$ (with upright typeface) to denote
the configuration projector onto and against the CI space ${\sf P}$, respectively; 
$\hat{{\sf P}} = \sum_{\bf I}^{\sf P} \vert {\bf I}\rangle\langle{\bf I}\vert$, and
$\hat{{\sf Q}} = \hat{1} - \hat{{\sf P}}$.
Equations~(\ref{eq:g-tdci}) and (\ref{eq:g-tdmo}) are the general EOMs
for CI coefficients and orbitals, respectively, valid for MCSCF
wavefunctions with arbitrary CI spaces {\sf P}.

Equation~(\ref{eq:g-tdmo}) suggests that the set of
orbital rotations $\{\hat{E}^\mu_\nu\}$ can be classified into the
following disjoint categories:
\begin{enumerate}
\item {\it Redundant}.
Both 
$\hat{E}^\mu_\nu \vert \Psi \rangle$ and
$\hat{E}^\nu_\mu \vert \Psi \rangle$ lie inside ${\sf P}$ or vanish.
In this case, Eq.~(\ref{eq:g-tdmo}) reduces to an identity 
(thus called {\it redundant}), and $X^\mu_\nu$ may be arbitrary
anti-Hermitian matrix elements \cite{Caillat2005PRA}:
\begin{eqnarray}
\label{eq:x_red}
X^\mu_\nu = \langle \phi_\mu \vert \hat{\theta}(t) \vert \phi_\nu
\rangle, \hspace{.5em} \hat{\theta}^\dagger(t) = -\hat{\theta}(t).
\end{eqnarray}

\item {\it Non-redundant uncoupled}.
{\color{black}
At least one of 
$\hat{E}^\mu_\nu \vert \Psi \rangle$ and
$\hat{E}^\nu_\mu \vert \Psi \rangle$ do not vanish, and 
$\hat{E}^\mu_\nu \vert \Psi \rangle$ and
$\hat{E}^\nu_\mu \vert \Psi \rangle$ lie, if non-vanishing, outside
${\sf P}$.
}
Such rotations do not contribute to the CI equations,
Eq.~(\ref{eq:g-tdci}) (thus called {\it uncoupled}).
In this case, Eq.~(\ref{eq:g-tdmo}) reduces to a simpler expression
\cite{Sato:2013}, 
\begin{eqnarray}\label{eq:rmat_inter}
&
\langle \Psi \vert \left[\hat{E}^\mu_\nu, \hat{E}^\gamma_\lambda\right] \vert
\Psi \rangle X^\gamma_\lambda = -{\rm i}
\langle \Psi \vert \left[\hat{E}^\mu_\nu, \hat{H}\right] \vert
\Psi \rangle.
&
\end{eqnarray} 

\item {\it Non-redundant coupled}.
Either
$\hat{E}^{\mu}_{\nu} \vert \Psi \rangle$ or
$\hat{E}^{\nu}_{\mu} \vert \Psi \rangle$ 
lies across ${\sf P}$ and ${\sf Q}$. 
Such rotations do contribute to both the CI and orbital
EOMs (thus called {\it coupled}). In this case, one needs to
directly work with Eq.~(\ref{eq:g-tdmo}).

\end{enumerate}

\subsection{Analyses of orbital rotations in the ORMAS wavefunction
\label{subsec:ormas_anal}}
In what follows, we use orbital indices
$\{i,j,k\}$ for core ($\mathcal{C}$),
$\{t,u,v,w,x,y\}$ for active ($\mathcal{A}$), $\{p,q,r,s\}$ for occupied
($\mathcal{P}$), $\{a,b,c\}$ for virtual ($\mathcal{Q}$), and
$\{\mu,\nu,\lambda,\gamma,\delta\}$ for general ($\mathcal{H}$)
orbitals. The whole set of orbital rotations within the $\mathcal{H}$
space is categorized as follows:
\begin{eqnarray}
\left\{\hat{E}^\mu_\nu\right\} = 
\left\{\hat{E}^i_j, \hat{E}^i_t, \hat{E}^t_i, \hat{E}^t_u, \hat{E}^p_a, \hat{E}^a_p, \hat{E}^a_b\right\}.
\end{eqnarray}
Reference~\cite{Sato:2013} identifies
the core-core and virtual-virtual rotations
$\{\hat{E}^i_j,\hat{E}^a_b\}$ as redundant, and 
core-active and occupied-virtual rotations
$\{\hat{E}^t_i,\hat{E}^i_t,\hat{E}^a_p,\hat{E}^p_a\}$ as
non-redundant uncoupled, for the CASSCF wavefunction.
This conclusion is valid for general CI space ${\sf P}$, since
\textcolor{black}{the}
derivation of Ref.~\cite{Sato:2013} makes no use of the internal
structure of the active space (complete or non-complete) for these parameters.
Furthermore, for the same reason, the final expression of relevant time derivative terms
$\{X^t_i,X^i_t,X^a_p\}$ of TD-CASSCF method applies
to general MCSCF wavefunctions with no modifications.

Left unexplored above is the active-active rotations
$\{\hat{E}^t_u\}$, which we analyze as follows.
First, active intra-group rotations $\{E^t_u; \phi_t,\phi_u \in \mathcal{A}_g\}$
are redundant, since such rotations do not change the occupation
distribution, and the expansion of Eq.~(\ref{eq:cip_dist}) is complete
for a given distribution;
for every $|{\bf I}\rangle \in {\sf P}(\bm{d}) \subset {\sf P}$,
$\hat{E}^t_u|{\bf I}\rangle \in {\sf P}(\bm{d}) \subset {\sf P}$,
thus $\hat{E}^t_u|\Psi\rangle \in {\sf P}$.
Next, active inter-group rotations $\{E^t_u; \phi_t \in
\mathcal{A}_g, \phi_u \in \mathcal{A}_{g^\prime}, g \ne
g^\prime\}$ are, 
in general, non-redundant coupled. This is understood by
considering the simplest example of Fig.~\ref{fig:ormas}~(b) with, e.g., single and
double excitations from the first into the second subgroup included.
In this example, the CI space is given by
$
{\sf P} = {\sf P}(6,0) + {\sf P}(5,1) + {\sf P}(4,2).
$
Then if $|{\bf I}\rangle \in {\sf P}(6,0)$, then
$\hat{E}^t_u|{\bf I}\rangle \in {\sf P}(5,1) \subset {\sf P}$ where $\phi_u$ and
$\phi_t$ belong to the first and second subgroups, respectively. However for the same rotation, 
if $|{\bf I}\rangle \in {\sf P}(4,2)$, then $\hat{E}^t_u|{\bf
I}\rangle \in {\sf P}(3,3) \subset {\sf Q}$, 
thus $\hat{E}^t_u|\Psi\rangle$ lies across ${\sf P}$ and ${\sf Q}$.
See Ref.~\cite{Miyagi:2013} for a similar discussion.

\subsection{Final expression of TD-ORMAS orbital equations of motion\label{subsec:td-ormas_mo}}
As pointed out in the previous subsection, the equations for the CI
coefficients and orbitals except the terms 
$\{X^t_u\}$ are independent of the active space structure.
Thus we write down the final expression of EOMs by referring to the
TD-CASSCF formulae \cite{Sato:2013}, first for orbitals in
this subsection and for CI coefficients in the next subsection, with
active-active terms $\{{X^t_u}\}$ left unspecified until Sec.~\ref{subsec:td-ormas_aa}.
The orbital EOMs are given by
{\color{black}
\begin{eqnarray}\label{eq:tdmo}
\vert\dot{\phi}_p\rangle &=&
- {\rm i}\hat{Q} \hat{F}_p \vert\phi_p\rangle
+ \sum_q \vert \phi_q \rangle X^q_p,
\end{eqnarray}
}
where $\hat{Q}$ is the orbital projector onto the $\mathcal{Q}$ space;
\begin{eqnarray}\label{eq:projq} 
\hat{Q} \equiv \sum_a |\phi_a\rangle\langle\phi_a| =
\hat{1} - \sum_p |\phi_p\rangle\langle\phi_p|,
\end{eqnarray}
which prevents the explicit use of virtual
orbitals \cite{Caillat2005PRA}, and 
\begin{subequations}\label{eq:fock}
\begin{eqnarray}
\label{eq:fockc}
{\color{black}
\hat{F}_i
} |\phi_i\rangle &=& 
\hat{f} |\phi_i\rangle + 
\sum_{tu} D^t_u \hat{G}^u_t |\phi_i\rangle, \\
\label{eq:focka}
\hat{F}_t \vert\phi_t\rangle &=&
\hat{f}|\phi_t\rangle + \sum_{uvwx}
\hat{W}^v_w \vert\phi_u\rangle P^{uw}_{xv} \left(D^{-1}\right)^x_t,
\end{eqnarray}
\end{subequations}
where 
$D^t_u \equiv \langle\Psi_\textrm{A}|\hat{E}^u_t
|\Psi_\textrm{A}\rangle$ and 
$P^{tv}_{uw} \equiv
\langle\Psi_\textrm{A}|\hat{E}^{uw}_{tv}|\Psi_\textrm{A}\rangle$ are 
one- and two-electron reduced density matrix (RDM) elements,
respectively, defined within the active space, and
\textcolor{black}{
\begin{eqnarray}
\label{eq:cfock}
\hat{f} |\phi_p\rangle &=& 
\hat{h} |\phi_p\rangle +
2 \sum_j \hat{G}^j_j |\phi_p\rangle, \\
\label{eq:afock}
\hat{G}^p_q|\phi_r\rangle &=& \hat{W}^p_q|\phi_r\rangle -
\frac{1}{2}\hat{W}^p_r|\phi_q\rangle, \\
\label{eq:meanfield}
W^p_q(\bm{r}_1) &=&
\int d{\color{black}\bm{r}_2} 
\frac{\phi^*_p({\color{black}\bm{r}_2}) \phi_q({\color{black}\bm{r}_2})}
{|\bm{r}_1 - \bm{r}_2|}.
\end{eqnarray}
}

The core-active term $X^t_i$ is given \cite{Sato:2013} by the
solution of the following matrix equation:
\begin{eqnarray}
\label{eq:tdmo_ca}
(2\delta^t_u - D^t_u) X^u_i &=& -{\rm i} B^t_i,
\end{eqnarray}
and $X^i_t = -X^{t*}_i$, where
\begin{eqnarray}
\label{eq:bmat_ca}
B^t_i &\equiv& \langle\Psi|\left[\hat{E}^i_t,\hat{H}\right]|\Psi\rangle =
2F^t_i - D^t_u F^{i*}_u, \\
\label{eq:bmat_aa}
B^t_u &\equiv& \langle\Psi|\left[\hat{E}^u_t,\hat{H}\right]|\Psi\rangle =
\sum_v \left(F^t_v D^v_u - D^t_v F^{u*}_v\right),
\end{eqnarray}
are the so called Brillouin matrix elements used in the stationary MCSCF
methods, and
{\color{black}
\begin{eqnarray}
\label{eq:fmat_gbt}
F^p_q &=& \langle\phi_p|\cdot\hat{F}_q|\phi_q\rangle.
\end{eqnarray}
}

With no core orbitals, the core fock operator of Eq.~(\ref{eq:cfock})
reduces to the bare one-electron operator $\hat{h}$, whereas if core
orbitals are classified into frozen (fixed in time) and dynamical
(allowed to vary in time) core orbitals \cite{Sato:2013},
the range of core indices $i,j$ should be restricted to dynamical cores in
all equations in this section, with the operator $\hat{h}$ in
Eq.~(\ref{eq:cfock}) replaced with $\hat{h}^\textrm{FC}$ 
given by
\begin{eqnarray}
\label{eq:cfock_frozen}
\hat{h}^\textrm{FC}(t) &=& 
\hat{h}(t) + 2 \sum_k^\textrm{FC} \hat{G}^k_k(0),
\end{eqnarray}
where the summation $k$ is restricted within the frozen-core orbitals.
Equation~(\ref{eq:cfock_frozen}) emphasizes the fact that the (direct
and exchange) two-electron contributions from the frozen-core electrons,
$\hat{G}^k_k(0) \equiv \hat{G}^k_k(t=0)$ are time-independent.

\subsection{Final expression of TD-ORMAS CI equations of motion\label{subsec:td-ormas_ci}}
The CI equation is given as follows \cite{Sato:2013};
\begin{eqnarray}\label{eq:tdci}
\dot{C}_{\bf I} = -{\rm i}
\langle{\bf I} \vert \hat{H}_\textrm{A} - E_\textrm{A}\hat{1} \vert \Psi_\textrm{A} \rangle -
\langle{\bf I} \vert \hat{X} \vert \Psi_\textrm{A} \rangle,
\end{eqnarray}
\begin{eqnarray}
\label{eq:hama}
\hat{H}_\textrm{A} &=&
\sum_{tu} f^t_u \hat{E}^t_u + \frac{1}{2} \sum_{tuvw} g^{tv}_{uw}
\hat{E}^{tv}_{uw},
\end{eqnarray}
where $\hat{1}$ is a unit operator, $E_\textrm{A} \equiv
\langle\Psi_\textrm{A}|\hat{H}_\textrm{A}|\Psi_\textrm{A}\rangle$,
and 
\begin{eqnarray}
\label{eq:fmat}
f^t_u &=& \langle\phi_t| \cdot \hat{f}|\phi_u\rangle, \\
\label{eq:gmat}
g^{tv}_{uw} &=& \langle\phi_t| \cdot \hat{W}^v_w|\phi_u\rangle.
\end{eqnarray}
In Eq.~(\ref{eq:tdci}), we make, without loss of generality, a
particular phase choice so that $\langle \Psi \vert \dot{\Psi} \rangle =
0$. Another, more common choice of the phase
${\rm i}\langle\Psi|\dot{\Psi}\rangle = \langle\Psi|\hat{H}|\Psi\rangle$ 
replaces the operator $\hat{H}_\textrm{A} - E_\textrm{A}\hat{1}$ in
Eq.~(\ref{eq:tdci}) with
$\hat{H}_\textrm{A} + E_\textrm{C}\hat{1}$ where $E_\textrm{C} = 2\sum_j f^j_j$.
These approaches are mathematically equivalent, but the former improves
the stability of both real and imaginary propagations \cite{Sato:2013}.
The separation of the core wavefunction in Eq.~(\ref{eq:mcscf_2q})
allows to formulate the CI equation as the effective
$N_\textrm{A}$-electron problem [Eq.~(\ref{eq:tdci})], 
rather than that of the total $N$ electrons [Eq.~(\ref{eq:g-tdci})].

\subsection{Active inter-group contributions\label{subsec:td-ormas_aa}}
Now we turn to the active inter-group rotations $\{E^t_u\}$
to derive the equation for $\{X^t_u\}$.
Let us re-emphasize that 
Eqs.~(\ref{eq:tdmo}) and (\ref{eq:tdci}) 
are valid irrespective of the active space structure.
The equation to be solved for $\{X^t_u\}$, derived in this subsection, 
thus finalizes our derivation of the TD-ORMAS method.

Although Eq.~(\ref{eq:g-tdmo}) is useful 
for the formal discussion as made in Sec.~\ref{subsec:tdvp}, 
it does not fully take into account the anti-Hermiticity of matrices
$\Delta$ and $X$. Thus, instead of starting from Eq.~(\ref{eq:g-tdmo}), 
we directly work with real and imaginary parts of $\Delta$ and $X$;
\begin{eqnarray}
\label{eq:vorb}
\Delta^t_u &=& \Delta^{\rm R}_{tu} + {\rm i} \Delta^{\rm I}_{tu}, \\
\label{eq:dorb}
X^t_u &=& X^{\rm R}_{tu} + {\rm i} X^{\rm I}_{tu}.
\end{eqnarray}
Here 
$\Delta^{\rm R}, X^{\rm R}$ are anti-symmetric, and
$\Delta^{\rm I}, X^{\rm I}$ are symmetric. The active inter-group
parts of operators $\hat{\Delta}$ and $\hat{X}$ are now expressed as
\begin{eqnarray}
\label{eq:var_acc}
\hat{\Delta} &=& {\sum_{t>u}}^\prime \left(
\Delta^{\rm R}_{tu} \hat{E}^{-}_{tu} + {\rm i} 
\Delta^{\rm I}_{tu} \hat{E}^{+}_{tu}\right), \\
\label{eq:der_acc}
\hat{X} &=& {\sum_{t>u}}^\prime \left(
X^{\rm R}_{tu} \hat{E}^{-}_{tu} + {\rm i} 
X^{\rm I}_{tu} \hat{E}^{+}_{tu}\right),
\end{eqnarray}
where $\hat{E}^{\mp}_{tu} = \hat{E}^t_u \mp \hat{E}^u_t$.
The primed summations in these equations are taken over active
inter-group rotations, which amounts to $N_\textrm{rot}$ nonequivalent
rotations with 
\begin{eqnarray}\label{eq:nrot}
N_\textrm{rot} = \sum_{g>g^\prime}^G n_g n_{g^\prime}.
\end{eqnarray}
Inserting Eqs.~(\ref{eq:var_acc}) and
(\ref{eq:der_acc}) into Eq.~(\ref{eq:varder}), and requiring that
$\delta S$ of Eq.~(\ref{eq:ds}) vanishes for $\Delta^{\rm R}_{tu}$ and
$\Delta^{\rm I}_{tu}$ separately, after straightforward rearranging of terms,
we have 
\begin{subequations}\label{eq:tdmo_aa}
\begin{eqnarray}
{\sum_{v>w}}^\prime \left(
A^{--}_{tu,vw} X^{\rm R}_{vw} + A^{-+}_{tu,vw} X^{\rm I}_{vw}
\right) &=& b^-_{tu}, \\
{\sum_{v>w}}^\prime \left(
A^{+-}_{tu,vw} X^{\rm R}_{vw} + A^{++}_{tu,vw} X^{\rm I}_{vw}
\right) &=& b^+_{tu},
\end{eqnarray}
\end{subequations}
where
\begin{eqnarray}
\label{eq:amat1}
A^{\mp\mp}_{tu,vw} &=& {\pm} {\rm Im} 
\langle\Psi_\textrm{A}| \hat{E}^{\mp}_{tu}\hat{\sf{Q}}\hat{E}^{\mp}_{vw} |\Psi_\textrm{A}\rangle, \\
\label{eq:amat2}
A^{\mp\pm}_{tu,vw} &=& {\pm }{\rm Re} 
\langle\Psi_\textrm{A}| \hat{E}^{\mp}_{tu}\hat{\sf{Q}}\hat{E}^{\pm}_{vw} |\Psi_\textrm{A}\rangle, \\
\label{eq:bvec1}
b^-_{tu} &=& -{\rm Re}
\langle\Psi_\textrm{A}| \hat{E}^-_{tu} \hat{\sf{Q}} \hat{H}_\textrm{A} |\Psi_\textrm{A}\rangle, \\
\label{eq:bvec2}
b^+_{tu} &=& +{\rm Im}
\langle\Psi_\textrm{A}| \hat{E}^+_{tu} \hat{\sf{Q}} \hat{H}_\textrm{A} |\Psi_\textrm{A}\rangle.
\end{eqnarray}

Equation~(\ref{eq:tdmo_aa}) is the desired formulae for the active
inter-group contributions to the orbital time derivative.
Although our main focus is on the use of the ORMAS model, 
this equation is in fact valid for the general MCSCF wavefunction with
arbitrary CI spaces {\sf P} since it is equivalent to
Eq.~(\ref{eq:g-tdmo}), the general equation.
However, without a systematic construction of CI spaces, in general,
all pairs of active orbitals with $N_\textrm{rot} =
n_\textrm{A}(n_\textrm{A}+1)/2$, instead of Eq.~(\ref{eq:nrot}), have to
be included, with no control of (non-) redundancy of active-active
rotations. The advantage of the TD-ORMAS method is that it can
limit the application of Eq.~(\ref{eq:tdmo_aa}) to 
non-redundant, inter-group pairs only.
This improves both the efficiency and stability of the temporal propagation.

\section{Implementation\label{sec:implementation}}
This section describes our implementation of the TD-ORMAS method.
Input parameters required are the number of core ($n_\textrm{C}$) and
active ($n_\textrm{A}$) orbitals, the number of active subgroups $G$,
and the size and occupation boundaries of each subgroup; 
\begin{eqnarray}
\bm{n} &=&
 (n_1,n_2,\cdot\cdot\cdot,n_G), \\
\bm{N}_\textrm{min} &=&
 (N^\textrm{min}_1,N^\textrm{min}_2,\cdot\cdot\cdot,N^\textrm{min}_G), \\
\bm{N}_\textrm{max} &=&
 (N^\textrm{max}_1,N^\textrm{max}_2,\cdot\cdot\cdot,N^\textrm{max}_G).
\end{eqnarray}
{\color{black} Our code first checks if the given input is a sensible
one. Thereafter,} the possible set of occupation distributions $\bm{d}$ and
the information of the CI space are automatically generated using the
algorithm of Ref.~\cite{Ivanic:2003a}. 

Note that (some of) the inter-group rotations may turn out to be
redundant, e.g., due to the symmetry. Our code removes such
redundant rotations from Eqs.~(\ref{eq:tdmo_aa}), if detected in advance. 
It can also happen that some inter-group rotations are identified
as non-redundant uncoupled, e.g., when a subgroup $\mathcal{A}_g$ has a fixed occupation
$N^\textrm{min}_g = N^\textrm{max}_g$, which makes all rotations involving
an orbital in $\mathcal{A}_g$ and the other outside $\mathcal{A}_g$ non-redundant uncoupled. 
For such cases, we still use the general equation
[Eq.~(\ref{eq:tdmo_aa})], since the reduction to simpler expression
[Eq.~(\ref{eq:rmat_inter})] leads to no significant computational gains.

Any type of propagators require the evaluation of time derivatives of
variables $\{\dot{C}_I, \dot{\phi}_p\}$ from the set of variables $\{C_I, \phi_p\}$.
This proceeds as follows;
\begin{enumerate}
\renewcommand{\labelenumi}{(\arabic{enumi})}
\item \label{enum:flow_rdm}
      Compute the active space RDMs $\left\{D^t_u, P^{tv}_{uw}\right\}$ from
      the current CI coefficients. We use the algorithm of
      Ref.~\cite{Panin:1996} to efficiently handle the ``coupling
      coefficients'' $\langle{\bf I}|\hat{E}^t_u\hat{E}^v_w|{\bf
      I}^\prime\rangle$ in non-complete CI spaces. 
\item Compute $\mathcal{Q}$-space contributions to the orbital derivatives
      [the first term of Eq.~(\ref{eq:tdmo})]
      from the current orbitals and RDMs obtained in the \textcolor{black}{step} (1).
      {\color{black}
      This is done by first evaluating the one-electron operator
      acting on occupied orbitals $\hat{h}|\phi_p\rangle$ and 
      meanfield operators $\hat{W}^p_q$ defined in
      Eq.~(\ref{eq:meanfield}),
      from which $\hat{f}|\phi_p\rangle$ and $\hat{G}^t_u|\phi_i\rangle$
      are evaluated according to Eqs.~(\ref{eq:cfock}) and
      (\ref{eq:afock}), respectively, and the right-hand sides of
      Eqs.~(\ref{eq:fockc}) and (\ref{eq:focka}) are accumulated.
      }
      To avoid the possible (near) singularity of inverse 1RDM in 
      Eq.~(\ref{eq:focka}), the eigenvalues $\left\{d_t\right\}$
      of $D$ are regularized as $1/d_t \rightarrow d_t/(d_t^2 +
      \delta^2)$, where $\delta$ is a small positive number.

\item Compute the one- and two-electron Hamiltonian elements, $f^t_u$
      and $g^{tv}_{uw}$ entering Eq.~(\ref{eq:hama}) by
      performing the inner products of Eqs.~(\ref{eq:fmat}) and
      (\ref{eq:gmat}), using $\hat{f}|\phi_t\rangle$ and $\hat{W}^p_q$
      obtained in the \textcolor{black}{step} (2).
\item Compute the Brillouin matrix elements $B^p_q$ 
      entering Eqs.~(\ref{eq:tdmo_ca}) and (\ref{eq:bbar_1rdm})
      by performing the inner products of
      Eq.~(\ref{eq:fmat_gbt}),
      using $\hat{F}_p|\phi_p\rangle$
      obtained in the \textcolor{black}{step} (2).
\item Compute the direct Hamiltonian contribution to the CI derivative [the
      first term of Eq.~(\ref{eq:tdci})] from the matrix elements
      $f^t_u$, $g^{tv}_{uw}$ evaluated in the
      \textcolor{black}{step} (3) and the 
      current CI coefficients. The algorithm of Ref.~\cite{Panin:1996}
      is also used here.
\item Compute the core-active contributions to the orbital derivative
      $\{X^t_i, X^i_t\}$ by solving Eq.~(\ref{eq:tdmo_ca}).
      The regularization method given above is applied to the 
      matrix $2{\bf 1} - D$, where ${\bf 1}$ is a $n_\textrm{A} \times
      n_\textrm{A}$ identity matrix. 
\item Compute the active inter-group contributions to the orbital
      derivative $\{X^t_u\}$. First, the matrix elements of
      Eqs.~(\ref{eq:amat1})-(\ref{eq:bvec2}) are evaluated by the method
      described in Appendix~\ref{app:amat_bvec}. Then 
      Eq.~(\ref{eq:tdmo_aa}) is formulated as a real-valued matrix equation
      with the dimension $2N_\textrm{rot}$;
      \begin{eqnarray}\label{eq:amat}
      {\sf A} \bm{x} = \bm{b},
      \end{eqnarray}
      where ${\sf A} \equiv ((A^{--},A^{+-})^{\rm t},
      (A^{-+},A^{++})^{\rm t})$, 
      $\bm{b} \equiv (b^{-},b^{+})^{\rm t}$, 
      and $\bm{x}$ is the solution vector whose
      first and last $N_\textrm{rot}$ elements being the real and imaginary
      parts of $X$, respectively. Here the matrix ${\sf A}$ is
      anti-symmetric for a real time derivative
      [Eqs.~(\ref{eq:amat1}) and (\ref{eq:amat2})], and symmetric for an
      imaginary time derivative [Eqs.~(\ref{eq:amat1_imag}) and
      (\ref{eq:amat2_imag})].
      This equation is solved by the singular value decomposition of the
      coefficient matrix {\sf A}, with its singular values being regularized
      by the same procedure as used in the
      \textcolor{black}{steps} (2) and (6). 

\item Add in the completed $\mathcal{P}$-space orbital derivative matrix 
      $\left\{X^p_q\right\}$ both to the orbital equation
      [Eqs.~(\ref{eq:tdmo})] and to the CI equation [Eq.~(\ref{eq:tdci})]. 
\end{enumerate}

For the full MCTDHF method, \textcolor{black}{steps} (1)-(3), (5) [and (8) if
$\hat{\theta} \neq 0$ in Eq.~(\ref{eq:x_red})] complete
the evaluation of time derivatives. The TD-CASSCF method
requires \textcolor{black}{steps} (1)-(6) and (8). For more general cases,
all \textcolor{black}{steps} have to be executed. See Ref.~\cite{Sato:2013}
for more detailed explanation of these \textcolor{black}{steps} except (7). 
The efficient algorithm given in Appendix~\ref{app:amat_bvec} allows
the \textcolor{black}{step} (7) to be performed with a very small computational cost.

The above-described procedures are used both for real-time propagations
and imaginary-time propagations (to obtain the stationary
state). For the latter case, each propagation is followed by the
normalization of CI coefficients and the Schmidt orthonormalization of
orbitals.
One should be careful in transforming the real-time EOMs
into the imaginary-time ones in the case of non-complete CI spaces.
Appendix~\ref{app:imag} explicitly gives the equations appropriate
for the imaginary-time propagation.

\section{Applications\label{sec:applications}}
\begin{figure}[!b]
\centering
\includegraphics[width=.90\linewidth,clip]{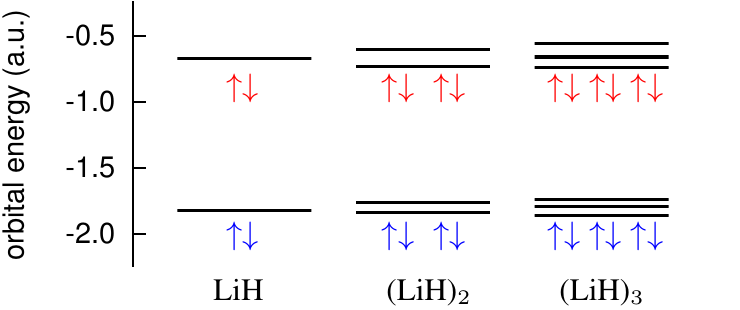}
\caption{\label{fig:orbene}Hartree-Fock orbital energy
levels of 1D-LiH clusters. 
Numerical values obtained for (LiH)$_3$ are
-1.860, -1.794, -1.742, -0.747, -0.661, and -0.565. 
Those for LiH (-1.824 and -0.674) and (LiH)$_2$ (-1.848, -1.767, -0.728, and
 -0.599) are taken from Ref.~\cite{Sato:2013}.
Red and blue arrows indicate electrons
occupying the weakly and deeply bound orbitals, respectively.
}%
\end{figure}

In this section, we apply the TD-ORMAS method to the one-dimensional
(1D) model systems.
The 1D multielectron models have served as convenient but reliable
testing ground for assessing new theoretical methods
\cite{Pindzola:1991,Pindzola:1997,Dahlen:2001,Caillat2005PRA,Nguyen:2006,Sato:2013,Miyagi:2013,Sato:2014,Miyagi:2014a,Miyagi:2014b}.
By doing this we demonstrate the flexibility of the
ORMAS framework, and discuss (dis)advantages of various 
options of active spaces.
The 1D model Hamiltonian for $N$ electrons in the potential of $M$
fixed nuclei interacting with an external laser electric field $E(t)$ is taken as
\begin{eqnarray}
\label{eq:1d-ham}
H &=& \sum_i^N \left\{-\frac{1}{2}\frac{\partial^2}{\partial x^2_i}
- \sum_a^M \frac{Z_a}{\sqrt{(x_i - X_a)^2 + c}}
- E(t)x_i
\right\} \nonumber \\
&+& \sum_{i>j}^N \frac{1}{\sqrt{(x_i-x_j)^2 + d}},
\end{eqnarray} 
where $x_i$ is the position of the $i$-th electron, $\{X_a\}$ and
$\{Z_a\}$ are the positions and charges of nuclei, and $c=0.5$
and $d=1$ \cite{Sato:2013} adjust the soft Coulomb operators of electron-nuclear and
electron-electron interactions, respectively. 
The electron-laser interaction is included within the dipole approximation 
and in the length gauge.
Note that the result is gauge invariant \cite{Sato:2013, Miyagi:2014b}. 
The redundant orbital rotations of Eq.~(\ref{eq:x_red}) are fixed as
$\hat{\theta} = 0$, and the regularization parameter $\delta$ introduced
in Sec.~\ref{sec:implementation} is taken to be sufficiently small
(typically $\delta = 10^{-10}$). The orbital EOMs are discretized on
equidistant grid points with spacing $\Delta x = 0.4$ and box size $|x|
< 600$. Further computational details are the same as in Ref.~\cite{Sato:2013}.

Specifically, we investigate 1D lithium hydride (LiH) cluster models,
1D-(LiH)$_m$ \cite{Sato:2013}, with $m = 1, 2, 3$.
We consider the collinear configuration, LiH-LiH-$\cdot\cdot\cdot$
(nuclear charges $3131\cdot\cdot\cdot$), with
interatomic LiH distance 2.3 and intermolecular H-Li distance 3.5 as
optimized for (LiH)$_2$ \cite{Sato:2013}.
Figure~\ref{fig:orbene} shows the ground-state Hartree-Fock orbital
energies. As shown in the figure, the
electronic structure of (LiH)$_m$ consists of $m$ tightly
bound orbitals (with $2m$ electrons) and $m$ weakly bound
orbitals (with $2m$ electrons). This is the consequence of
the strong bonding interaction within LiH and the weak intermolecular
interaction between LiH molecules. 
We have previously found the following observations \cite{Sato:2013}
for $m = 1,2$;
\begin{enumerate}
\renewcommand{\labelenumi}{\arabic{enumi})}
\item The lowest $n_\textrm{C}=m$ orbitals can be treated as core in a
      very good approximation: The TD-CASSCF method, with only the upper
      $N_\textrm{A}=2m$ electrons treated as active, closely reproduces
      the full MCTDHF results.
\item At least $n_\textrm{A}=2N_\textrm{A}$ orbitals are required for $N_\textrm{A}$
      active electrons to obtain the convergent results for, e.g., the
      temporal evolution of the dipole moment and ionization yields in
      the presence of an intense laser field.
\end{enumerate}

As a preliminary to the present work, we confirmed that the latter 
conclusion is valid also for (LiH)$_3$; successive TD-CASSCF
calculations with $N_\textrm{A} = 6$ and increasing $n_\textrm{A}$
reached the convergence at $n_\textrm{A} = 12$ for
the above observables. We could not perform the full MCTDHF calculation
with $n = n_\textrm{C} + n_\textrm{A} = 15$ orbitals and
$N=N_\textrm{C}+N_\textrm{A}=12$ electrons due to
the large CI dimension (more than 25 million determinants) which exceeds
the capability of our present computational code. However, one
reasonably expects a similar accuracy for TD-CASSCF descriptions of
(LiH)$_3$ as for those of smaller systems. In this work, therefore, we use
the TD-CASSCF method with $N_\textrm{A} = 2m$ active electrons and
$n_\textrm{A} = 2N_\textrm{A}$ active orbitals, abbreviated by
CAS($2N_\textrm{A}$), as a standard.

\begin{table}[!t]
\caption{\label{tab:gsprop}
Ground state properties of 1D-(LiH)$_3$ model. The number of active
orbitals $n_\textrm{A}$, the number of determinants $N_\textrm{det}$, the
total energy $E$, and the dipole moment $\langle x \rangle$ are shown
for various methods.}
\begin{ruledtabular}
\begin{tabular}{rrrcc}
\multicolumn{1}{r}{Method}&
\multicolumn{1}{r}{$n_{\rm A}$} &
\multicolumn{1}{r}{$N_{\rm det}$} &
\multicolumn{1}{c}{$E$}&
\multicolumn{1}{c}{$\langle x \rangle$}\\
\hline
\multicolumn{5}{c}{HF reference CI wavefunctions} \\
\multicolumn{1}{r}{HF}     &  3 &          1 & -21.2125   & -3.128 \\
\multicolumn{1}{r}{+S}     &  6 &         19 & -21.2300   & -3.214 \\
\multicolumn{1}{r}{+SD}    & 12 &      1,000 & -21.2636   & -3.336 \\
\multicolumn{1}{r}{+SDT}   & 12 &      7,000 & -21.2647   & -3.352 \\
\multicolumn{1}{r}{+SDTQ}  & 12 &     23,200 & -21.2653   & -3.356 \\
\multicolumn{5}{c}{CAS(6) reference CI wavefunctions} \\
\multicolumn{1}{r}{CAS(6)} &  6 &        400 & -21.2540   & -3.335 \\
\multicolumn{1}{r}{+S}     & 12 &      4,000 & -21.26{\color{black}35}   & -3.3{\color{black}49} \\
\multicolumn{1}{r}{+SD}    & 12 &     15,700 & -21.2652   & -3.356 \\
\multicolumn{1}{r}{+SDT}   & 12 &     32,700 & -21.2653   & -3.356 \\
\multicolumn{5}{c}{RASCI wavefunctions} \\
\multicolumn{1}{r}{RAS($3,1$)}   & 12 &      2,082 & -21.26{\color{black}31}   & -3.34{\color{black}3} \\
\multicolumn{1}{r}{RAS($3,2$)}   & 12 &      5,340 & -21.2648   & -3.350 \\
\multicolumn{1}{r}{RAS($4,2$)}   & 12 &     11,955 & -21.2652   & -3.355 \\
\multicolumn{1}{r}{RAS($4,3$)}   & 12 &     20,455 & -21.2653   & -3.356 \\
\multicolumn{1}{r}{CAS(12)}& 12 &     48,400 & -21.2653   & -3.356 \\
\end{tabular}
\end{ruledtabular}


\end{table}
Table~\ref{tab:gsprop} shows the ground state properties of 1D-(LiH)$_3$
obtained with various methods, grouped 
according to the type of underlying CI spaces discussed in Sec.~\ref{subsec:ormas}.
The rigorous definitions of these methods are given below in
Secs.~\ref{subsec:rhf+}-\ref{subsec:ras}. At this point, we mention that
these classes of methods provide different series of approximations, 
whose accuracy can be improved systematically until final
convergence to the CAS($2N_\textrm{A}$) description.
Table~\ref{tab:gsprop} demonstrates such a systematic improvement for
each class of methods, where the total energy and the dipole moment of
the ground state are converged to the CAS($2N_\textrm{A}$)
values to {\color{black} four and three decimal places, respectively}. The
higher accuracy is achieved at the expense of higher computational cost,
as shown in the steep increase of the CI dimension in
table~\ref{tab:gsprop}. The question is then how fast, with respect to 
the level of approximation
within each class of methods, the adequate
accuracy is obtained for a given physical problem.

In the following three subsections, we address this question for the
description of intense-field driven multielectron dynamics.
For this purpose, we consider the temporal evolution of the dipole
moment as a basic measure of the accuracy. The high-harmonic generation (HHG) spectrum is
investigated in Sec.~\ref{subsec:hhg} to see the performance of methods
to predict an experimentally relevant observable.
Appendix~\ref{app:ormas} includes numerical test of more complex ORMAS
wavefunction than those assessed in this section, addressing a difficulty
encountered in propagating such a complex wavefunction.
We consider a laser field of the following form;
\begin{eqnarray}\label{eq:laser}
E(t) = E_0 \sin(\omega_0 t) \sin^2\left(\pi \frac{t}{\tau}\right),
\hspace{.5em} 0 \leq t \leq \tau.
\end{eqnarray}
with laser parameters corresponding to \textcolor{black}{a
wavelength of 750 nm (period $T \approx 2.5$ fs), 
a peak intensity of 4$\times$10$^{14}$ W/cm$^2$,
and a duration of three optical cycles ($\tau = 3T \approx 7.5$ fs)}.

\subsection{Hartree-Fock reference CI wavefunctions\label{subsec:rhf+}}
First we assess the HF+X methods. The ORMAS parameters for the (LiH)$_m$
models are set as $G = 2$,
$\bm{n} = (N_\textrm{A}/2, 3N_\textrm{A}/2)$,
$\bm{N}_\textrm{min} = (L, 0)$, and
$\bm{N}_\textrm{max} = (N_\textrm{A}, L)$, with $N_\textrm{A} = 2m$,
generating following CI spaces;
\begin{subequations}\label{eq:rhf+}
\begin{eqnarray}
\label{eq:rhf+_lih}
{\sf P}_\textrm{LiH} &=&
\sum_{l=0}^L
\left[\phi_1\right]^{2-l}
\left[\phi_2\uuline{\phi_3\phi_4}\right]^l, \\
\label{eq:rhf+_lih2}
{\sf P}_\textrm{(LiH)$_2$} &=&
\sum_{l=0}^L
\left[\phi_1\phi_2\right]^{4-l}
\left[\phi_3\phi_4\uuline{\phi_5\textrm{--}\phi_8}\right]^l, \\
\label{eq:rhf+_lih3}
{\sf P}_\textrm{(LiH)$_3$} &=&
\sum_{l=0}^L
\left[\phi_1\phi_2\phi_3\right]^{6-l}
\left[\phi_4\phi_5\phi_6\uuline{\phi_7\textrm{--}\phi_{12}}\right]^l.
\end{eqnarray}
\end{subequations}
%
We exceptionally set $\bm{n} =
(N_\textrm{A}/2,N_\textrm{A}/2)$ for the HF+S method 
[doubly underlined orbitals are removed in Eqs.~(\ref{eq:rhf+})], 
for which only $n_\textrm{A} \leq N_\textrm{A}$ is meaningful \cite{Miyagi:2014a}.
This class of methods has been proposed
and assessed for 1D models of helium, beryllium, and carbon
atoms in Ref.~\cite{Miyagi:2014b}. These models
are similar to our 1D-(LiH)$_m$ models with $m = 1, 2, 3$ in
the sense that helium, beryllium, and carbon atoms are two, four, and six electron
systems, respectively. However, since the inner most orbitals of 1D beryllium and
carbon models are energetically far apart from the other orbitals
\cite{Miyagi:2014b},  
they indeed represent effective two (helium), two (beryllium), and four
(carbon) electron problems under the investigated laser parameters
\cite{Miyagi:2014b}.
In contrast, our 1D-(LiH)$_m$ models involve
{\it equally important} $N_\textrm{A} = 2m$ active electrons as shown in
Fig.~\ref{fig:orbene}, thus serve as more stringent test cases. 

\begin{figure}[!t]
\centering
\includegraphics[width=.90\linewidth,clip]{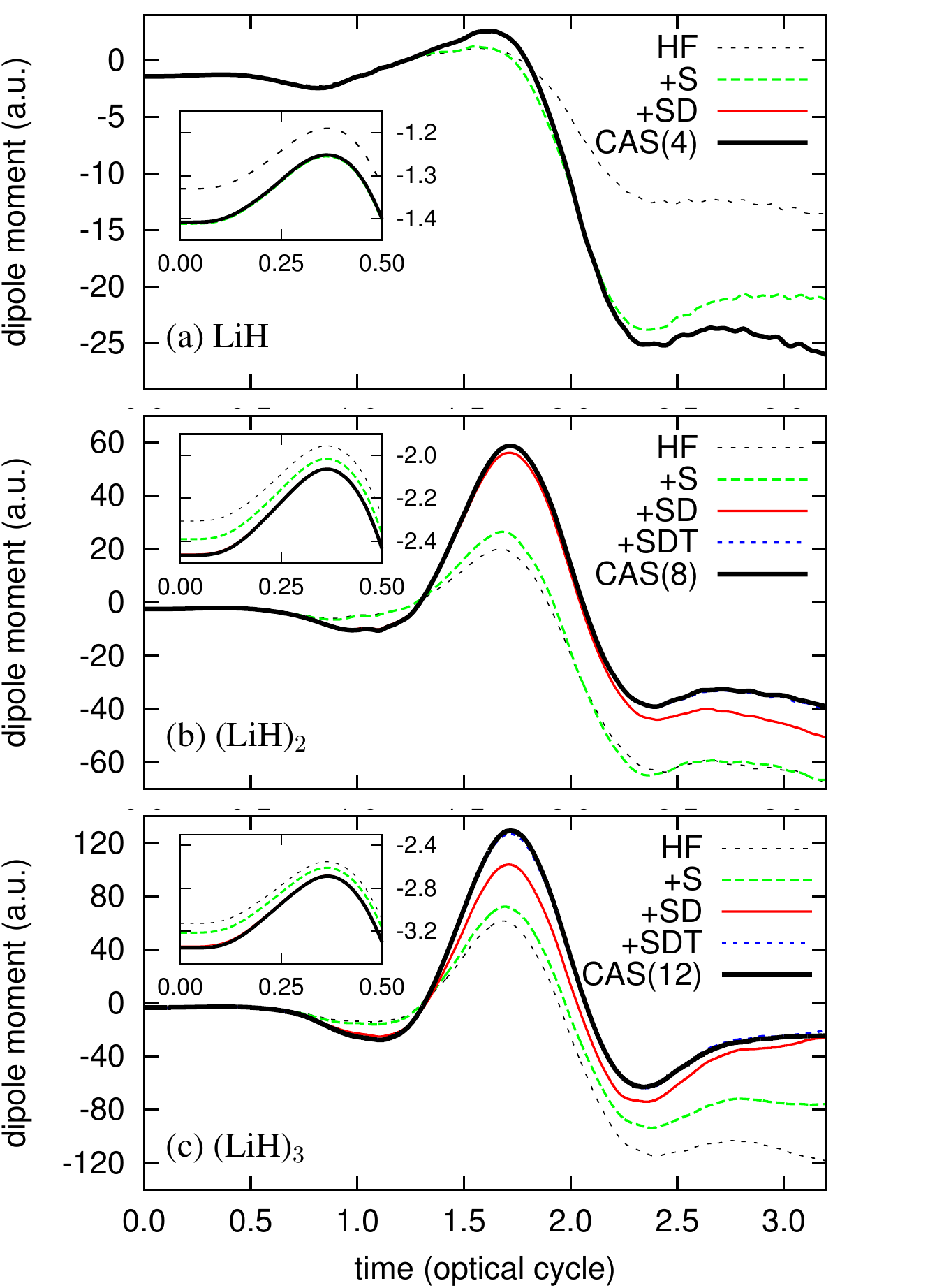}
\caption{\label{fig:dip_rhf+}The time evolution of the dipole moment of
(a) LiH, (b) (LiH)$_2$, and (c) (LiH)$_3$ models, computed with HF and
HF+X methods (with X=S, SD, SDT signifying $L$ = 1, 2, 3, respectively)
compared with the CAS($2N_\textrm{A}$) results.}%
\end{figure}

Figure~\ref{fig:dip_rhf+} shows the evolution of the dipole moment
computed with HF and HF+X methods. 
The HF method gives the dipole with large deviations from that of
CAS($2N_\textrm{A}$) for all $m = 1,2,3$.
The HF+X methods with $L \geq 2$ offer \textcolor{black}{a} substantially better description,
showing the steady convergence to the CAS($2N_\textrm{A}$)
description with increasing $L$.
%
However it should be noted that the convergence rate with respect to $L$ gets slower for
larger systems. For example, the HF+SD method ($L=2$) is
exact (equivalent to the CAS) for LiH with $N_\textrm{A}=2$ [Fig.~\ref{fig:dip_rhf+}~(a)], closely reproduces
the CAS($2N_\textrm{A}$) result for (LiH)$_2$ with $N_\textrm{A}=4$,
except a small deviation at the final stage of the pulse
[Fig.~\ref{fig:dip_rhf+}~(b)], but gives the dipole which noticeably
deviates from the CAS($2N_\textrm{A}$) result for (LiH)$_3$ with $N_\textrm{A} = 6$
[Fig.~\ref{fig:dip_rhf+}~(c)]. 
We also note that a larger value of $L$ is required to
properly describe dynamics than the static electronic structure;
the HF+SD describes the ground state of (LiH)$_3$ very well as seen in
the inset of Fig.~\ref{fig:dip_rhf+}~(c), 
but the accuracy gets deteriorated in the presence the
electron-laser interaction. Meanwhile, we observe that the HF+S method gives a
reasonably accurate result only for LiH, but brings no major improvement
over the HF description for larger systems. This is in contrast to
Ref.~\cite{Miyagi:2014a} which reported a good performance of this
method for 1D model atoms.

The distinct advantage of the HF+X method, with a fixed $L$, is the polynomial
scaling of the computational cost against $N_\textrm{A}$, 
as emphasized in Ref.~\cite{Miyagi:2014b}.
However as noted above, the accuracy of the HF+X with a fixed $L$ rapidly drops
for larger systems, and depends on the electronic structure in hand. 
The first difficulty (size dependence) is the consequence of the lack of the
size-extensivity \cite{Szabo:1996,Helgaker:2002}. The latter problem
(situation dependence) is related to the fundamental limitation of the
Hartree-Fock wavefunction; the closed-shell wavefunction cannot 
properly describe tunneling ionization process \cite{Sato:2013, Sato:2014}. The 
Hartree-Fock reference determinant, in the HF+X method, is no longer a
good starting point, demanding the inclusion of higher excitations
to describe the more delocalized wavefunction that arises during the
course of tunneling ionization.

\subsection{CAS($N_\textrm{A}$) reference CI wavefunctions\label{subsec:cas+}}
Next we consider the CAS($N_\textrm{A}$)+X methods. The ORMAS parameters
are identical to those of the HF+X methods except for the different
allocation of $n_\textrm{A} = 2N_\textrm{A}$ active orbitals as $\bm{n} =
(N_\textrm{A}, N_\textrm{A})$. The resultant CI spaces for (LiH)$_m$
models with $m = 2, 3$ read
\begin{subequations}
\label{eq:cas+}
\begin{eqnarray}
\label{eq:cas+_lih2}
{\sf P}_\textrm{(LiH)$_2$} &=&
\sum_{l=0}^L
\left[\phi_1\textrm{--}\phi_4\right]^{4-l}
\left[\phi_5\textrm{--}\phi_8\right]^l, \\
\label{eq:cas+_lih3}
{\sf P}_\textrm{(LiH)$_3$} &=&
\sum_{l=0}^L
\left[\phi_1\textrm{--}\phi_6\right]^{6-l}
\left[\phi_7\textrm{--}\phi_{12}\right]^l,
\end{eqnarray}
\end{subequations}
This class of methods has been proposed in Ref.~\cite{Miyagi:2014b}, 
but not numerically investigated. 
We do not show results for LiH,
since no new approximations can be generated
in the case of $N_\textrm{A}=2$.
One expects that the CAS($N_\textrm{A}$)
reference serves as better starting point than the single HF reference
in the HF+X method, thus remedies the 
undesirable
system dependence of the accuracy of the latter approach.

\begin{figure}[!t]
\centering
\includegraphics[width=.90\linewidth,clip]{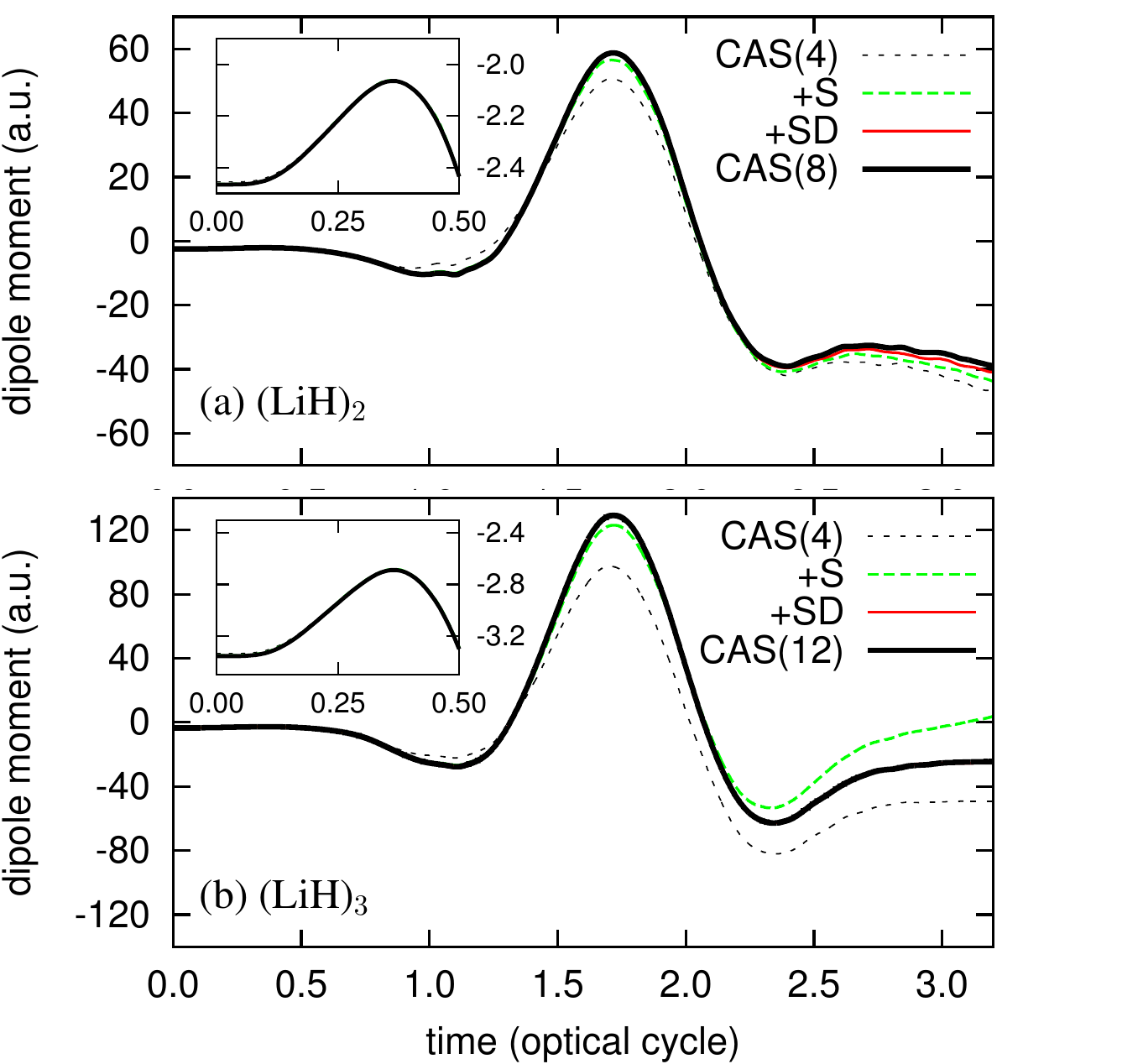}
\caption{\label{fig:dip_cas+}The time evolution of the dipole moment of
(a) (LiH)$_2$, and (b) (LiH)$_3$ models, computed with CAS($N_\textrm{A}$) and
CAS($N_\textrm{A}$)+X methods 
(with X=S, SD signifying $L$ = 1, 2, respectively)
compared with the CAS($2N_\textrm{A}$) results.}%
\end{figure}

This expectation is verified
in Fig.~\ref{fig:dip_cas+}, which compares the CAS($N_\textrm{A}$) and
CAS($N_\textrm{A}$)+X dipoles with the CAS($2N_\textrm{A}$) ones for
(LiH)$_2$ and (LiH)$_3$. 
As seen in the figure, both for (LiH)$_2$ and (LiH)$_3$, the
CAS($N_\textrm{A}$) dipole shows much better agreement with 
the CAS($2N_\textrm{A}$) result than does the HF one (Fig.~\ref{fig:dip_rhf+}), although the
large-amplitude oscillation during the second laser cycle is not
completely followed. The CAS($N_\textrm{A}$)+SD method gives the dipoles
with excellent agreement with those of CAS($2N_\textrm{A}$), and
even the CAS($N_\textrm{A}$)+S method also reproduces
the CAS($2N_\textrm{A}$) results surprisingly well.

The reduced system dependence,
thus the more uniform accuracy,
{\color{black}
which is not much affected by different system sizes
and different stages of the electron-laser interaction,
}
is achieved by
accounting for the most important part of the electron correlation with small CAS
expansion ($N_\textrm{A}$ orbitals for $N_\textrm{A}$ electrons), 
which enables the {\it remaining} correlation to be included
with low rank excitations. In the present case,
the first kind of correlation is the breakdown of the
closed-shell dominance during the course of tunneling ionization
\cite{Sato:2014}. This is analogous to the {\it static} correlation
involved in the bond breaking process \cite{Szabo:1996,Helgaker:2002}. 
Although the size-extensivity is still missing in the CAS($N_\textrm{A}$)+X method, the resultant error
is considerably reduced from that in the HF+X approach.
The drawback is the greater number of determinants involved compared to the
HF+X method with the same $L$, as shown in table~\ref{tab:gsprop} for
the (LiH)$_3$ case. The cost scales exponentially with respect to
$N_\textrm{A}$, as just so does the CAS($2N_\textrm{A}$) method.

\subsection{RAS CI wavefunctions\label{subsec:ras}}
To pursue further flexibility, we consider the RASCI space mentioned
in Sec.~\ref{subsec:ormas}. We set $G = 3$ and $\bm{n} =
(N_\textrm{A}/2, N_\textrm{A}/2, N_\textrm{A})$ with $N_\textrm{A} = 2m$.
The CI spaces for (LiH)$_2$ and (LiH)$_3$ can be written as
\begin{subequations}\label{eq:ras}
\begin{eqnarray}
\label{eq:ras_lih2}
{\sf P}_\textrm{(LiH)$_2$} &=&
\sum_{l_2=0}^{M_\textrm{elec}}\left\{\sum_{l_1=0}^{M_\textrm{hole}}
\left[\phi_1\phi_2\right]^{N_\textrm{A}-l_1}
\left[\phi_3\phi_4\right]^{l_1-l_2}\right\}
\left[\phi_5\textrm{--}\phi_8\right]^{l_2}, \nonumber \\ \\
\label{eq:ras_lih3}
{\sf P}_\textrm{(LiH)$_3$} &=&
\sum_{l_2=0}^{M_\textrm{elec}}\left\{\sum_{l_1=0}^{M_\textrm{hole}}
\left[\phi_1\textrm{--}\phi_3\right]^{N_\textrm{A}-l_1}
\left[\phi_4\textrm{--}\phi_6\right]^{l_1-l_2}\right\}
\left[\phi_7\textrm{--}\phi_{12}\right]^{l_2}. \nonumber \\
\end{eqnarray}
\end{subequations}
The factor within the braces in Eqs.~(\ref{eq:ras}), with $l_2 = 0$,
represents the CI space with the HF determinant ($l_1 = 0$) plus up to
$M_\textrm{hole}$-fold excitations ($l_1 > 0$) to the second subgroup.
It serves as the reference CI space, from which further excitations ($l_2
> 0$) to the third subgroup are to be included. 
Note that if $M_\textrm{hole} = M_\textrm{elec}$, Eqs.~(\ref{eq:ras}) reduce
to Eqs.~(\ref{eq:rhf+}), while if $M_\textrm{hole} = N_\textrm{A} >
M_\textrm{elec}$, Eqs.~(\ref{eq:ras}) are identical to
Eqs.~(\ref{eq:cas+}), with $L = M_\textrm{elec}$.
In this way, the present RAS scheme provides a flexible series of
approximations that includes the HF+X and CAS($N_\textrm{A}$)+X approaches
as special cases,
with two accuracy (cost) controlling parameters $M_\textrm{hole}$ and $M_\textrm{elec}$.
See Fig.~\ref{fig:ormas}~(d) for a pictorial understanding.

To estimate a reasonable value of $M_\textrm{hole}$, we performed
preliminary calculations (not shown) with $M_\textrm{elec} = 0$.
They correspond to the HF+X calculations
with the active space reduced by half from that in
Sec.~\ref{subsec:rhf+};
$\bm{n} = (N_\textrm{A}/2, N_\textrm{A}/2)$.
We have found that the (reduced) HF+SD and HF+SDT methods approximate the
CAS($N_\textrm{A}$) method quite well for (LiH)$_2$ and (LiH)$_3$,
respectively, exactly as seen for the twice larger active space in
Sec.~\ref{subsec:rhf+}. 
Thus we use $M_\textrm{hole} = N_\textrm{A}/2 = m$, or more, for (LiH)$_m$.
In addition, we consider two possibilities $M_\textrm{elec} = 1, 2$,
having the good performance of CAS($N_\textrm{A}$)+X methods with $L = 1,2$ in
mind. 
For brevity, the method based on Eqs.~(\ref{eq:ras}) is denoted as
RAS($M_\textrm{hole},M_\textrm{elec}$). 
The RAS($N_\textrm{A}/2,1$)
and RAS($N_\textrm{A}/2,2$) methods are further abbreviated as RAS1 and
RAS2, respectively, which aim for reduced-cost alternatives to
CAS($N_\textrm{A}$)+S and CAS($N_\textrm{A}$)+SD methods, respectively.

\begin{figure}[!t]
\centering
\includegraphics[width=.90\linewidth,clip]{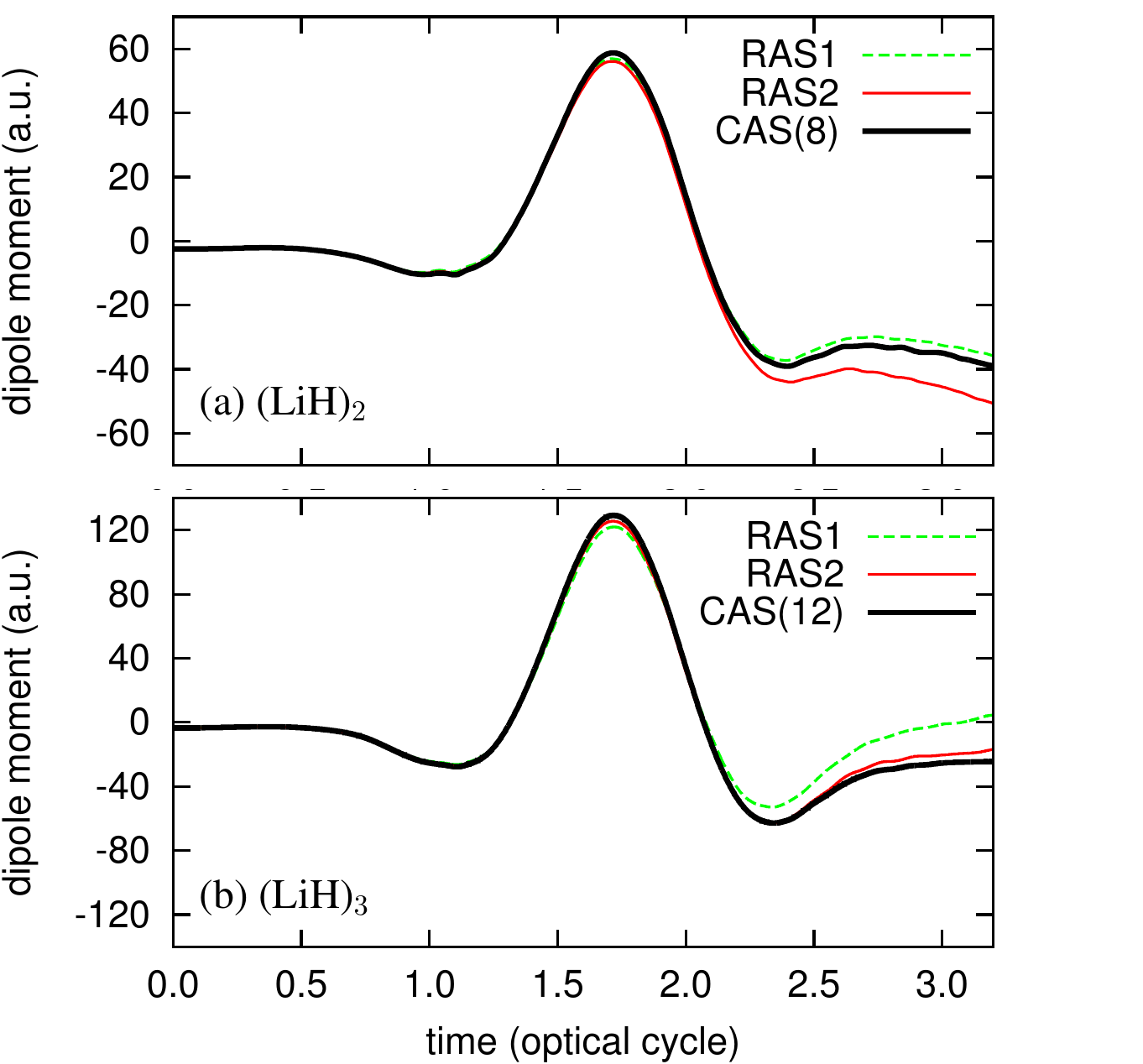}
\caption{\label{fig:dip_rasx}The time evolution of the dipole moment of
(a) (LiH)$_2$, and (b) (LiH)$_3$ models, computed with RAS1 and
RAS2 methods compared with the CAS($2N_\textrm{A}$) results.}%
\end{figure}
Figure~\ref{fig:dip_rasx} shows the dipole moment computed with the RAS1
and RAS2 methods for (LiH)$_2$ and (LiH)$_3$ models. As seen in the
figure, these methods closely reproduce the dipole evolution of the
CAS($2N_\textrm{A}$) method, including the global oscillation
at the center of the pulse. In closer look at the figure, the RAS1
dipole evolutions are found to be very similar to those of the
CAS($N_\textrm{A}$)+S method in Fig.~\ref{fig:dip_cas+}. 
The RAS2 result of (LiH)$_2$ is identical to that of HF+SD in
Fig.~\ref{fig:dip_rhf+}, as should be so since $M_\textrm{hole} = M_\textrm{elec}$. 
In whole, the performance of the
RAS methods is satisfactory especially when we notice the significant
reduction of the CI dimension as shown in table~\ref{tab:gsprop} for
(LiH)$_3$. We further confirmed (not shown)
that increasing $M_\textrm{hole}$ by one
($M_\textrm{hole}=N_\textrm{A}/2+1=m+1$) results in the dipole which is
indistinguishable, in the scale of the figure, from the corresponding
CAS($N_\textrm{A}$)+X one.
The high performance of the RAS schemes is attributed to the two-stage
approximations controlled by $M_\textrm{hole}$ and $M_\textrm{elec}$;
The reference CI space accounts for the (system-dependent) important part of the
correlation (with system-dependent $M_\textrm{hole}$), while 
the excited configurations are responsible for the remaining part (with
$M_\textrm{elec}$ typically up to doubles).

\subsection{High-harmonic generation spectrum\label{subsec:hhg}}
\begin{figure}[!b]
\centering
\includegraphics[width=.90\linewidth,clip]{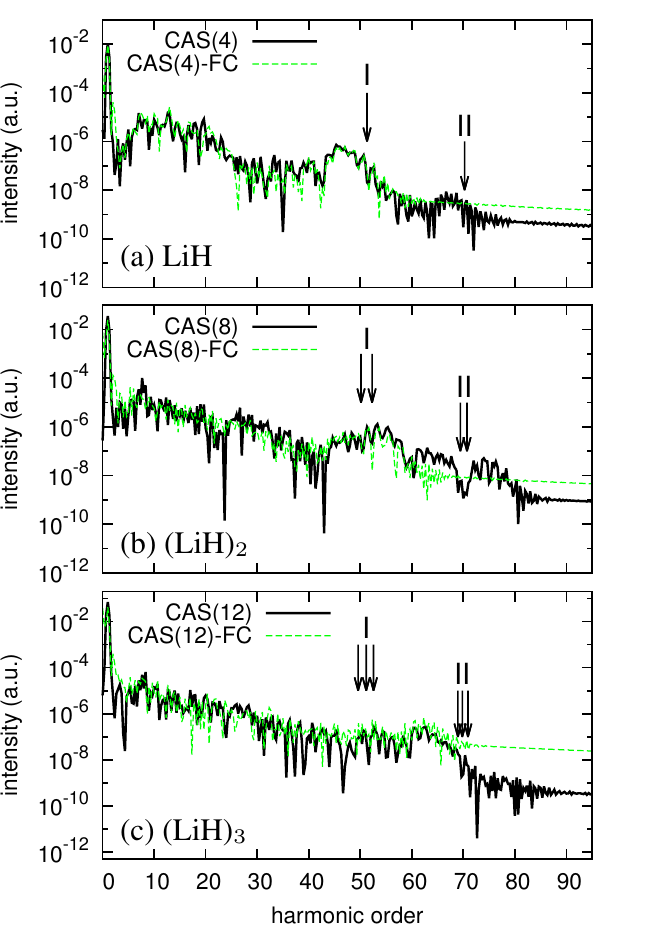}
\caption{\label{fig:hhg_cas}The HHG spectra of (a) LiH, (b) (LiH)$_2$,
and (c) (LiH)$_3$ models, 
{\color{black}
exposed to a laser pulse of the form Eq.~(\ref{eq:laser}) with a wavelength of 750
nm and an intensity of 4$\times$10$^{14}$ W/cm$^2$,
}
computed with CAS($2N_\textrm{A}$) methods.
The dynamical-core and frozen-core (``-FC'' appended) spectra are
compared. The three-step model prediction of cutoff positions are indicated
by arrows. See text for more details.}%
\end{figure}
\begin{widetext}

\begin{figure}[!t]
\centering
\includegraphics[width=1.\linewidth,clip]{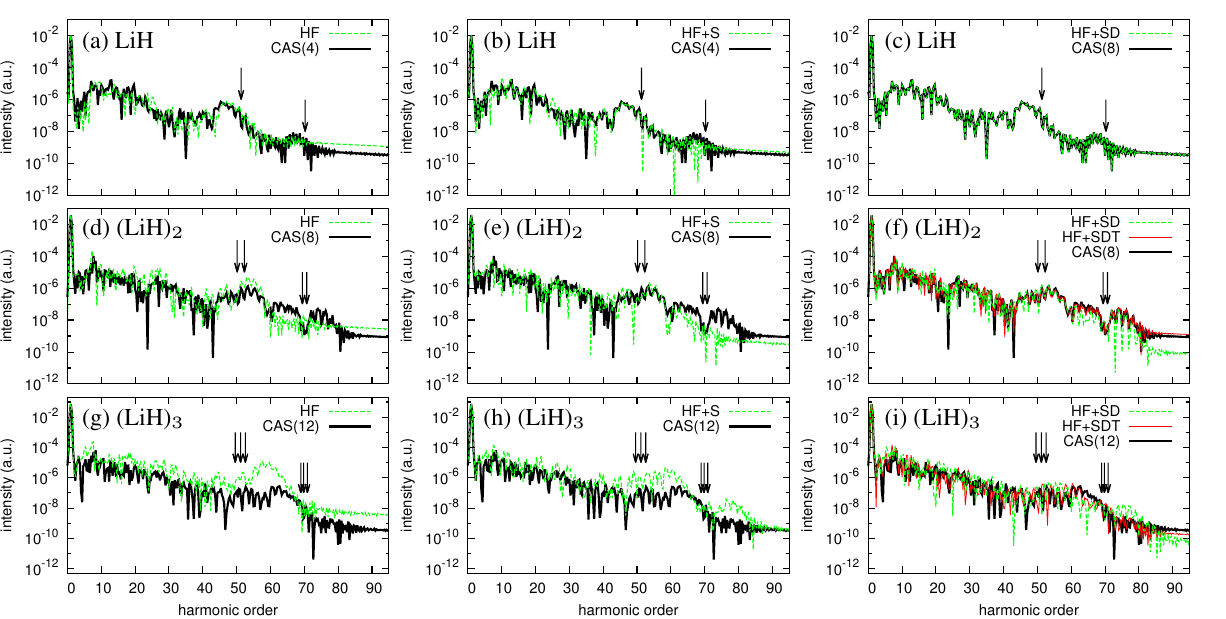}
\caption{\label{fig:hhg_rhf+}The HHG spectra of LiH (a-c), (LiH)$_2$ (d-f),
and (LiH)$_3$ (g-i) models, computed with HF and HF+X methods 
(with X=S, SD, SDT signifying $L$ = 1, 2, 3, respectively) compared
with CAS($2N_\textrm{A}$) spectra. Also see the caption of Fig.~\ref{fig:hhg_cas}.}%

\includegraphics[width=1.\linewidth,clip]{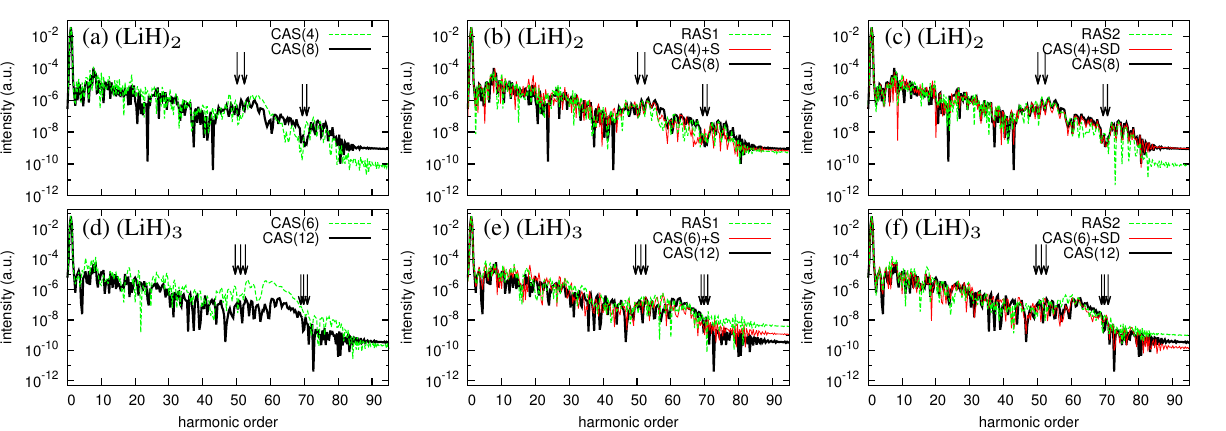}
\caption{\label{fig:hhg_cas+}The HHG spectra of (LiH)$_2$ (a-c),
and (LiH)$_3$ (d-f) models, computed with CAS($N_\textrm{A}$),
CAS($N_\textrm{A}$)+X (with X=S, SD signifying $L$ = 1, 2, respectively)
and RAS methods compared with CAS($2N_\textrm{A}$) spectra. Also see
the caption of Fig.~\ref{fig:hhg_cas}.} 
\end{figure}
\end{widetext}

Next we investigate HHG spectra.
The HHG spectrum is obtained by the Fourier
transform of the expectation value of the dipole acceleration evaluated
using the Ehrenfest expression \cite{Bandrauk:2013}.
Before entering the assessment of different methods, we comment on
the physical interpretation of the HHG spectra of (LiH)$_m$ models.
Figure~\ref{fig:hhg_cas} shows the HHG spectra computed with the
CAS(2$N_\textrm{A}$) methods.
Shown in the figure with downward arrows are the cutoff positions, calculated
based on the static Hartree-Fock-Koopmans picture; 
\begin{eqnarray}\label{eq:cutoff_3step}
\omega^\textrm{cutoff}_i = -\epsilon_i + 3.17 U_\textrm{p},
\end{eqnarray}
where $U_\textrm{p} \equiv E^2_0/4\omega^2_0$ is the ponderomotive energy, and $\epsilon_i$ is
the orbital energy depicted in Fig.~\ref{fig:orbene}. 
As seen in Fig.~\ref{fig:hhg_cas}~(a), the
computed HHG spectrum of LiH is characterized by the two-stage
cutoff structure, with the positions of the first and second cutoff being
well reproduced by Eq.~(\ref{eq:cutoff_3step}) with weakly (I) and deeply (II)
bound orbital energies, respectively.
The comparison of dynamical-core and frozen-core
treatments [denoted as CAS($2N_\textrm{A}$) and CAS($2N_\textrm{A}$)-FC, respectively]
reveals that the second cutoff originates from the core response, since it is absent in the frozen-core spectrum. 
This simple picture based on the independent particle model gets less valid in
larger systems [Fig.~\ref{fig:hhg_rhf+}~(b) and (c)]. 
The higher complexity of the spectra for larger systems is 
presumably attributed to the higher probability of multiple
ionizations and the increasing importance of the multichannel effect
with growing molecular size.
Further physical discussions of HHG spectra will be made elsewhere. 
Below we focus on how the various methods reproduce the CAS($2N_\textrm{A}$) spectra.

Figure~\ref{fig:hhg_rhf+} compares the HHG spectra 
computed with HF and HF+X methods with those of
CAS($2N_\textrm{A}$). 
As can be seen in Fig.~\ref{fig:hhg_rhf+}~(a), the HF method
already gives the HHG spectrum of LiH with a good agreement with that of
CAS(4), as opposed to the large deviation in the dipole moment [Fig.~\ref{fig:dip_rhf+}~(a)].
%
With the present laser setting, the high-harmonic emissions
are dominated by those during the second laser cycle ($T < t < 2T$).
As seen in Fig.~\ref{fig:dip_rhf+}~(a), for LiH, the electron motion within this
time region is restricted near the origin, which allows the HF method
for a qualitatively correct description.
In contrast, for (LiH)$_2$ and (LiH)$_3$, the HF method clearly
overestimates the spectral intensity below the first cutoff. 
%
%
The underestimation of tunneling ionization and first-order response, and
the overestimation of harmonic intensity for a high-intensity laser are
common faults of the TDHF method \cite{Pindzola:1991,Kulander:1992,Nguyen:2006,Sato:2013,Sato:2014}.
The argument for the performance of the HF+X methods goes parallel to
that for the dipole moment made in Sec.~\ref{subsec:rhf+};
Increasing $L$ steadily improves the description, but the accuracy with
a fixed $L$ gets poorer for larger systems. 
The HF+SDT method well reproduces the CAS($2N_\textrm{A}$) spectra up to (LiH)$_3$.
The HF+SD and CAS($2N_\textrm{A}$) spectra agree exactly and quite well for LiH and
(LiH)$_2$, respectively, but deviates noticeably for (LiH)$_3$.
The HF+S spectrum of (LiH)$_3$ is no better than that of the HF method.

The HHG spectra computed with CAS($N_\textrm{A}$), 
CAS($N_\textrm{A}$)+X, and RAS methods are shown in Fig.~\ref{fig:hhg_cas+}.
The performance of the CAS($N_\textrm{A}$) method
[Fig.~\ref{fig:hhg_cas+}~(a) and (d)] is found unsatisfactory, with
little improvement over the HF spectra [Fig.~\ref{fig:hhg_rhf+}~(d) and
(g)].
%
However, 
as in Figs.~\ref{fig:hhg_cas+}~(c) and (f), the CAS($N_\textrm{A}$)+SD
spectra show a quite good agreement with the CAS($2N_\textrm{A}$)
ones. This convinces us that the 
CAS($N_\textrm{A}$) description is indeed the adequate starting point
(qualitatively correct for the tunneling ionization event), on top of
which the remaining correlation effect is included effectively with
low rank excitations. It is encouraging that the CAS($N_\textrm{A}$)+S
method also gives rather accurate HHG spectra as shown in
Figs.~\ref{fig:hhg_cas+}~(b) and (e).
Finally, as in the case of dipole evolution, the RAS1 and RAS2 methods
perform similarly to the CAS($N_\textrm{A}$)+S and
CAS($N_\textrm{A}$)+SD methods, respectively, despite their
significantly reduced CI dimensions. 
Again, we confirmed (not shown) that the HHG spectrum of (LiH)$_3$ computed with the
RAS($4,2$) method [with $M_\textrm{hole}$ increased by one from the
RAS2 $\equiv$ RAS($3,2$) method] agrees almost perfectly with that of the
CAS($N_\textrm{A}$)+SD method.
The great advantage of the CAS($N_\textrm{A}$)+X and RAS
methods is that the accuracy with a fixed $L$ ($M_\textrm{elec}$) is not
lost for larger systems as badly as in the case of the HF+X method.

{\color{black}
\subsection{Analyses of computational cost\label{subsec:cost}}
\begin{table}[!b]
\caption{\label{tab:cpu}
Computational times for the 1D-(LiH)$_3$ model. The CPU times
(second) for the computational steps (1)+(5), (2)--(4), and (7)
of Sec.~\ref{sec:implementation} as well as the total CPU
time are shown for various methods. The total speed-up factor relative to the CAS(12)
method are also shown in parentheses. See text for more details.}
\begin{ruledtabular}
\begin{tabular}{rrrrrl}
\multicolumn{1}{r}{Method}&
\multicolumn{1}{r}{(1)+(5)} &
\multicolumn{1}{r}{(2)--(4)} &
\multicolumn{1}{r}{(7)} &
\multicolumn{2}{c}{Total} \\
\hline
\multicolumn{6}{c}{HF reference CI wavefunctions} \\
\multicolumn{1}{r}{HF+SD} &  196.6 & 688.2 &  3.7 &  898.2&(3.8) \\
\multicolumn{1}{r}{+SDT}  &  780.7 & 678.5 & 12.7 & 1481.1&(2.3) \\
\multicolumn{1}{r}{+SDTQ} & 1565.6 & 676.9 & 13.8 & 2265.9&(1.5) \\
\multicolumn{6}{c}{CAS(6) reference CI wavefunctions} \\
\multicolumn{1}{r}{CAS(6)+S} &  202.9 & 551.2 &  2.2 &  766.0&(4.5) \\
\multicolumn{1}{r}{+SD}      & 1041.5 & 680.0 &  9.9 & 1741.1&(2.0) \\
\multicolumn{1}{r}{+SDT}     & 1991.3 & 690.5 & 14.9 & 2706.4&(1.3) \\
\multicolumn{6}{c}{RASCI wavefunctions} \\
\multicolumn{1}{r}{RAS($3,1$)} &  157.2 &  556.4 &  2.8 &  726.6&(4.7) \\
\multicolumn{1}{r}{RAS($3,2$)} &  603.4 &  684.6 & 10.3 & 1308.5&(2.6) \\
\multicolumn{1}{r}{RAS($4,2$)} &  905.5 &  683.8 & 11.2 & 1610.9&(2.1) \\
\multicolumn{1}{r}{RAS($4,3$)} & 1456.2 &  686.9 & 15.8 & 2169.1&(1.6) \\
\multicolumn{1}{r}{CAS(12)}    & 2736.5 &  689.8 &  0.0 & 3434.9&      \\
\end{tabular}
\end{ruledtabular}
%
%
%

\end{table}
Finally we analyze the computational cost of the
TD-ORMAS methods using the simulation for (LiH)$_3$ as an example.
Table~\ref{tab:cpu} shows CPU times for propagating 1000 time
steps from the initial ground-state with various methods
using the algorithm described in Sec.~\ref{sec:implementation}, recorded on a
single Xeon processor with a clock frequency of 3.33 GHz.
It is encouraging that
the HF+SDT, CAS(6)+S, CAS(6)+SD,
RAS($3,1$), and RAS($3,2$) methods, 
which are reasonably accurate for (LiH)$_3$ as shown in
Secs.~\ref{subsec:rhf+}-\ref{subsec:hhg}, all 
reduce the total computational time compared to that of the
CAS(12) method, with relative speed-up factors 2.3, 4.5, 2.0, 4.7, and
2.6, respectively. 
Table~\ref{tab:cpu} also shows the
CPU times for the computational steps (1)+(5), (2)--(4), and (7)
described in
Sec.~\ref{sec:implementation}, separately. Their sum accounts for
more than 98\% of the total CPU time. As seen in the table,
the CPU times for steps (1)+(5) are reduced for the case of 
approximate methods
depending on the number of determinants $N_\textrm{det}$,
while those of steps (2)--(4) are roughly constant [except for CAS(6)+S and
RAS($3,1$) methods, mentioned shortly]. This is because the former step (RDMs and CI
derivatives) scales linearly with $N_\textrm{det}$, 
while the latter (orbital derivatives except for the active-active terms and
operator integrals) is independent of $N_\textrm{det}$,
and depends only on the number of orbitals and basis
functions or grid points \cite{Sato:2013}.

The CPU times for steps (2)--(4) of CAS(6)+S and RAS($3,1$)
methods are shorter than those of the other methods. This is due to the
higher sparsity of the 2RDM [originating, in turn, from the maximum occupancy 1
of the last orbital subgroup as shown in Fig.~\ref{fig:ormas}~(c) and
(d)], which reduces the cost for the second term of
Eq.~(\ref{eq:focka}). Also, the sparsity of the coupling coefficients of
these methods makes the steps (1)+(5) faster.
Consequently, the total CPU times of CAS(6)+S and RAS($3,1$)
methods are shorter than that of the HF+SD method despite their larger
$N_\textrm{det}$ as shown in table~\ref{tab:gsprop}. 
Finally, the step (7), which is unique to the method with
a non-complete CI space, is found to occupy less
than 1\% of the total CPU time, highlighting the high efficiency of the
algorithm given in Appendix~\ref{app:amat_bvec}.
As a whole, the flexibility of the TD-ORMAS method and its optimal
implementation enable computational cost reduction
without significant loss of accuracy.
}

\section{Summary\label{sec:summary}}

A new time-dependent multiconfiguration method is developed
based on the ORMAS scheme to construct non-complete CI spaces. The 
TD-ORMAS method attains further flexibility on top of the 
previously developed TD-CASSCF method \cite{Sato:2013}
by the subdivision of active orbitals into an
arbitrary number of subgroups and the occupation restriction posed for each subgroup of
orbitals. The equations of motion 
for the CI coefficients and orbital functions in the TD-ORMAS method,
derived based on the time-dependent variational principle, are
shown to be formally identical to those of the TD-CASSCF method, except
for the non-vanishing active inter-group terms of orbital time
derivatives. An efficient algorithm is devised to solve for the
inter-group contributions, circumventing the costly evaluation of the 
three-particle reduced density matrix. 
The core wavefunction is explicitly separated from the
active CI space [Eq.~(\ref{eq:mcscf_2q})], transforming the
original $N$-electron CI equation [Eq.~(\ref{eq:g-tdci})] to that of
$N_\textrm{A}$ active electrons [Eq.~(\ref{eq:tdci})].
The implementation of the TD-ORMAS method is described in depth,
allowing existent MCTDHF codes to be readily adapted to the TD-ORMAS method. 

Out of a variety of methods that fall within the TD-ORMAS framework, 
several representative classes of methods are studied in detail; 
the HF+X, CAS($N_\textrm{A}$)+X, and RAS methods.
Note that the present RAS method is the straightforward
time-dependent version of
the stationary RASSCF method, differently from the ``TD-RASSCF'' method
of Ref.~\cite{Miyagi:2014b} (See Sec.~\ref{subsec:ormas}).
All the investigated approaches provide a systematic series of
approximations that converge to the TD-CASSCF description, but at
different rates with respect to the level of approximation [the
value(s) of $L$, or $M_\textrm{hole}$ and $M_\textrm{elec}$]. 
Among these methods, the present numerical analyses highlight the 
RAS method (encompassing the former two as special cases) as the most
cost effective one, 
which allows the separate calibrations for the reference CI space (by varying
$M_\textrm{hole}$) and for the further excitations from the reference
(by varying $M_\textrm{elec}$), thus enabling more flexible convergence
studies and applications with a reliable accuracy. 
We plan to make further assessment of above-mentioned
and  other problem-specific TD-ORMAS methods
based on three-dimensional implementation. This article has worked out
the theoretical issues regarding the use of non-complete CI spaces, and
provides a solid ground for more realistic applications.


\begin{acknowledgments}
We thank Dr. H. Miyagi and Dr. L. B. Madsen in Aarhus University for discussions.
This research is supported in part by Grant-in-Aid for Scientific
Research (No.~23750007, 23656043, 23104708, 25286064, 26390076, and
26600111) from the Ministry of Education, Culture, Sports, Science and
Technology (MEXT) of Japan, and also by Advanced Photon Science Alliance
(APSA) project commissioned by MEXT. This research is also partially
supported by the Center of Innovation Program from Japan Science and
Technology Agency, JST.
\end{acknowledgments}

\appendix

\section{Evaluation of matrix elements of
  Eqs.~(\ref{eq:amat1})-(\ref{eq:bvec2})\label{app:amat_bvec}} 
To compute the matrix elements of
Eqs.~(\ref{eq:amat1})-(\ref{eq:bvec2}),
we first evaluate the following quantities;
\begin{eqnarray}
\label{eq:pbar}
\bar{P}^{uw}_{tv} &=& \langle\Psi_\textrm{A}|\hat{E}^t_u\hat{\sf Q}\hat{E}^v_w|\Psi_\textrm{A}\rangle, \\
\label{eq:bbar}
\bar{B}^u_t &=& \langle\Psi_\textrm{A}| \hat{E}^t_u\hat{\sf Q}\hat{H}_\textrm{A} -
\hat{H}_\textrm{A}\hat{\sf Q}\hat{E}^t_u |\Psi_\textrm{A}\rangle,
\end{eqnarray}
from which ${\sf A}$ and $\bm{b}$ are easily obtained.
First, the tensor $\bar{P}$ is obtained as a by-product in computing the
2RDM, since
\begin{eqnarray}\label{eq:pbar_2rdm}
P^{uw}_{tv} &=& \langle\Psi_\textrm{A}| E^t_uE^v_w - E^t_w\delta^v_u
|\Psi_\textrm{A}\rangle \nonumber \\ &=&
\langle\Psi_\textrm{A}| E^t_u(\hat{\sf {P}} + \hat{\sf {Q}})E^v_w - E^t_w\delta^v_u
|\Psi_\textrm{A}\rangle \nonumber \\ &=&
\langle\Psi_\textrm{A}| E^t_u\hat{\sf {P}}E^v_w|\Psi_\textrm{A}\rangle
+ \bar{P}^{uw}_{tv} - D^w_t\delta^u_v,
\end{eqnarray}
where the identities $\hat{E}^{tv}_{uw} = \hat{E}^t_u\hat{E}^v_w -
\hat{E}^t_w\delta^v_u$ and ${\sf \hat{P}} + {\sf \hat{Q}} = \hat{1}$
are used. Next, the $\bar{B}$ matrix
is computed as follows;
\begin{eqnarray}
\label{eq:bbar_1rdm}
\bar{B}^u_t &=& 
\langle\Psi_\textrm{A}|
\hat{E}^t_u(\hat{1} - \hat{\sf P})\hat{H}_\textrm{A} -
\hat{H}_\textrm{A}(\hat{1} - \hat{\sf P})\hat{E}^t_u
|\Psi_\textrm{A}\rangle \nonumber \\ &=&
\langle\Psi_\textrm{A}| \left[\hat{E}^t_u, \hat{H}_\textrm{A}\right] |\Psi_\textrm{A}\rangle -
\langle\Psi_\textrm{A}|
\hat{E}^t_u\hat{\sf P}\hat{H}_\textrm{A} -
\hat{H}_\textrm{A}\hat{\sf P}\hat{E}^t_u
|\Psi_\textrm{A}\rangle \nonumber \\ &=&
B^u_t - \left(D^{\prime u}_t - D^{\prime t*}_u\right),
\end{eqnarray}
where Eq.~(\ref{eq:bmat_aa}) is used, 
and $D^{\prime u}_t \equiv
\langle\Psi_\textrm{A}|\hat{E}^t_u|\Psi^\prime_\textrm{A}\rangle$, with 
\begin{eqnarray}
\label{eq:psibar}
|\Psi^\prime_\textrm{A}\rangle = \sum_{\bf I}^{\sf P} |{\bf I}\rangle C^\prime_{\bf I}, \hspace{.5em}
C^\prime_{\bf I} = \langle{\bf I}| \hat{H}_\textrm{A} |\Psi_\textrm{A}\rangle.
\end{eqnarray}
The transformed coefficient 
$C^\prime_{\bf I}$ in Eq.~(\ref{eq:psibar}) is a part of the CI
derivative~(\ref{eq:tdci}) obtained beforehand in the
\textcolor{black}{step} (5) of 
Sec.~\ref{sec:implementation}, and 
the 1RDM like matrix $D^\prime$ is easily
computed with a cost typically an order of magnitude smaller than that of 2RDM. 
As a consequence, $\bar{P}$ and $\bar{B}$ (therefore
${\sf A}$ and $\bm{b}$) can be obtained with a small additional
effort on top of all the other operations. 

Reference~\cite{Miyagi:2014b} took a different approach, involving the
explicit computation of a part of the third-order reduced density matrix,
\begin{eqnarray}\label{eq:3rdm-q}
\zeta^{uwy}_{tvx} \equiv 
\langle\Psi_\textrm{A}|
\hat{E}^t_u\hat{\sf Q}\hat{E}^{vx}_{wy}
|\Psi_\textrm{A}\rangle,
\end{eqnarray}
to evaluate the two electron contributions to Eq.~(\ref{eq:bbar}) as
\begin{eqnarray}\label{eq:bvec_2e}
\langle\Psi_\textrm{A}| \hat{E}^t_u\hat{\sf Q}\hat{H}_\textrm{A} 
|\Psi_\textrm{A}\rangle \leftarrow \frac{1}{2}\sum_{vwxy}
g^{vx}_{wy} \zeta^{uwy}_{tvx}.
\end{eqnarray}
This constitutes a sever computational bottleneck, thus hampers the
inclusion of high rank excitations ($L > 3$) in the case of large
active spaces.
The present algorithm [Eqs.~(\ref{eq:bbar_1rdm})-(\ref{eq:psibar})]
removes this bottleneck.

\section{Active inter-group contributions for RASSCF wavefunction\label{app:tdmo_aa_ras}}
Equation~(\ref{eq:tdmo_aa}) can be transformed into a simpler form in
case of the RASSCF wavefunction \cite{Olsen:1988}.
As noted in Ref.~\cite{Olsen:1988} for the RASCI space, while upward {\it excitations}
$E^t_u|\Psi_\textrm{A}\rangle$ ($t > u$) can create states lying 
across ${\sf P}$ and ${\sf Q}$ spaces, all {\it deexcited}
configurations $E^u_t|\Psi_\textrm{A}\rangle$ ($t > u$) belong to the
${\sf P}$ space or vanish.
Using this fact in Eqs.~(\ref{eq:amat1})-(\ref{eq:bvec2}) leads
\begin{eqnarray}\label{eq:tdmo_aa_ras}
{\sum_{v>w}}^\prime {\sf A}^{\rm RAS}_{tu,vw} X_{vw} = \bm{b}^{\rm RAS}_{tu},
\end{eqnarray}
where
\begin{eqnarray}
{\sf A}^{\rm RAS}_{tu,vw} &\equiv& \langle\Psi_\textrm{A}|\hat{E}^u_t\hat{\sf Q}\hat{E}^v_w|\Psi_\textrm{A}\rangle, \\
\bm{b}^{\rm RAS}_{tu} &\equiv& \langle\Psi_\textrm{A}|\hat{E}^u_t\hat{\sf
Q}\hat{H}_\textrm{A}|\Psi_\textrm{A}\rangle.
\end{eqnarray}
This is identical to the equation used in Ref.~\cite{Miyagi:2014b}.
Since the matrix equation~(\ref{eq:tdmo_aa_ras}) has the same dimension as
the general equation~(\ref{eq:tdmo_aa}) (solution vector consists of two
times $N_{\rm rot}$ real values), we chose to always solve
Eq.~(\ref{eq:tdmo_aa}).

\section{Imaginary time propagation\label{app:imag}}
Appropriate equations for the imaginary time propagation
can be derived from the action
integral~(\ref{eq:s}) defined across the pure imaginary time axis $t =
-{\rm i}\tau$ with a real variable $\tau$.
Noting the following dual correspondence;
{\color{black}
\begin{eqnarray}\label{eq:dual}
|\frac{\partial \Psi}{\partial t}\rangle = {\rm i}
|\frac{\partial \Psi}{\partial \tau}\rangle \leftrightarrow
\langle\frac{\partial \Psi}{\partial t}| = -{\rm i}
\langle\frac{\partial \Psi}{\partial \tau}|,
\end{eqnarray}
}
the imaginary time counterpart of Eq.~(\ref{eq:ds}) is obtained as
{\color{black}
\begin{eqnarray}\label{eq:ds_imag}
\delta S = \delta \langle\Psi|\hat{H}|\Psi\rangle
{\color{black}+}
\left(
\langle\delta\Psi|\frac{\partial\Psi}{\partial \tau}\rangle +
\langle\frac{\partial\Psi}{\partial \tau}|\delta\Psi\rangle
\right). \nonumber \\
\end{eqnarray}
}
Based on this expression, formally the same equations are derived for
the CI derivative [Eq.~(\ref{eq:tdci})], and $\mathcal{Q}$-space and
core-active contributions 
[Eqs.~(\ref{eq:tdmo}) and (\ref{eq:tdmo_ca})] to the
orbital derivative, except for a replacement $-{\rm i} \rightarrow -1$
in the first term of these equations. 
One should take into account the sign difference of the third term of
Eqs.~(\ref{eq:ds}) and (\ref{eq:ds_imag}) for active inter-group rotations,
in general.
This results in the equation identical to Eq.~(\ref{eq:tdmo_aa}) with
the coefficient matrix of Eqs.~(\ref{eq:amat1}) and (\ref{eq:amat2})
redefined as follows;
\begin{eqnarray}
\label{eq:amat1_imag}
{\sf A}^{\mp\mp}_{tu,vw} &=& {\pm} {\rm Re} 
\langle\Psi_\textrm{A}| \hat{E}^{\mp}_{tu}\hat{\sf{Q}}\hat{E}^{\mp}_{vw} |\Psi_\textrm{A}\rangle, \\
\label{eq:amat2_imag}
{\sf A}^{\mp\pm}_{tu,vw} &=& {\pm }{\rm Im} 
\langle\Psi_\textrm{A}| \hat{E}^{\mp}_{tu}\hat{\sf{Q}}\hat{E}^{\pm}_{vw} |\Psi_\textrm{A}\rangle,
\end{eqnarray}
constituting the real symmetric linear system of equations.
If both CI coefficients and orbitals are represented by real numbers (as
for the non-degenerate ground state without external magnetic field),
variations and time derivatives of orbitals are parameterized
only with real anti-symmetric part of Eqs.~(\ref{eq:vorb}) and
(\ref{eq:dorb}), leading a matrix equation with half the dimension of
Eq.~(\ref{eq:tdmo_aa});
\begin{eqnarray}
{\sum_{v>w}}^\prime\langle\Psi_\textrm{A}|\hat{E}^{-}_{tu}\hat{\sf
 Q}\hat{E}^{-}_{vw}|\Psi_\textrm{A}\rangle
X^{\rm R}_{vw}
= -\langle\Psi_\textrm{A}|\hat{E}^{-}_{tu}\hat{\sf
Q}\hat{H}_\textrm{A}|\Psi_\textrm{A}\rangle. \nonumber \\
\end{eqnarray}

\section{Test of more complex ORMAS wavefunction\label{app:ormas}}
All the simulations adopting the non-complete CI spaces presented in
Sec.~\ref{sec:applications} have no stability problem; the integration
of the EOMs is found to be as stable as that of the CAS method.
Here we give an example of the difficult case. 
We set $G = 3$, $\bm{n} = (1, 1, 6)$, $\bm{N}_\textrm{min} = (1, 1, 0)$,
and $\bm{N}_\textrm{max} = (2, 2, 2)$, generating the following CI space
for (LiH)$_2$:
\begin{eqnarray}\label{eq:ormas}
{\sf P} =
\sum_{l_1 = 0}^{1} \sum_{l_2 = 0}^{1}
\left[\phi_1\right]^{2-l_1}
\left[\phi_2\right]^{2-l_2}
\left[\phi_3\textrm{--}\phi_8\right]^{l_1+l_2}.
\end{eqnarray}

\begin{figure}[!b]
\centering
\includegraphics[width=.90\linewidth,clip]{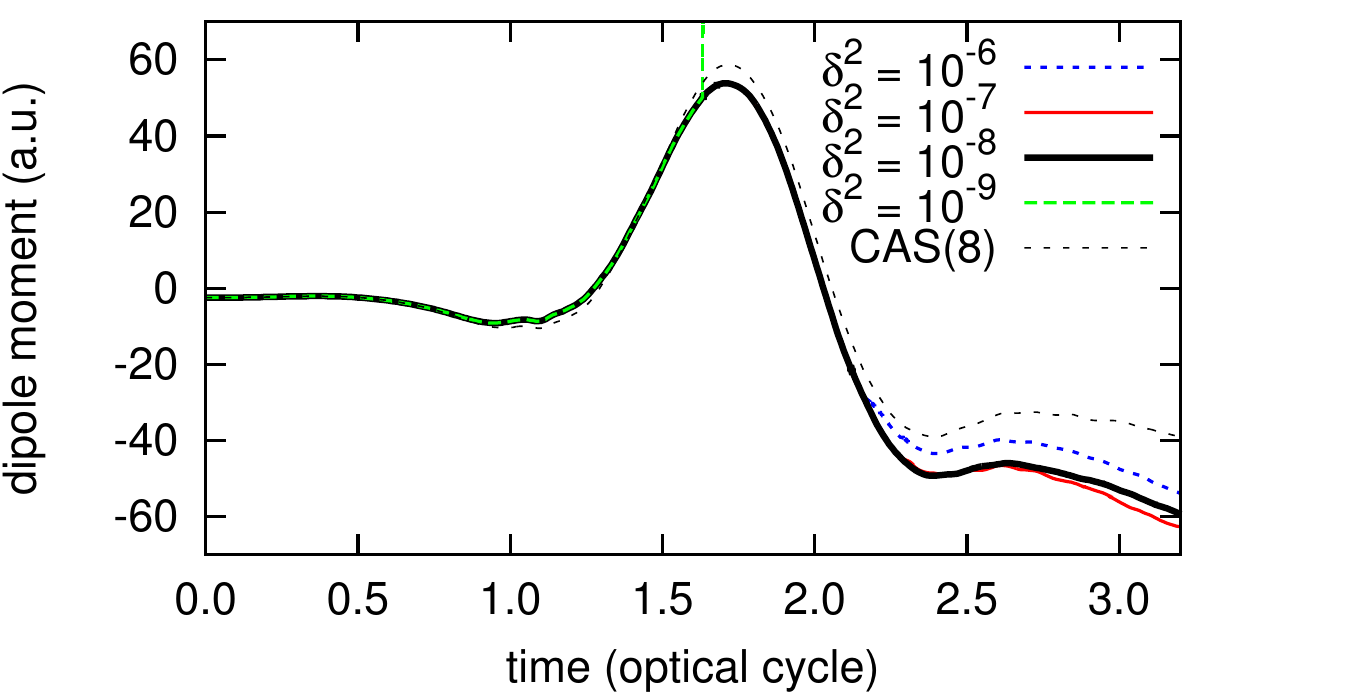}
\caption{\label{fig:dip_ormx}The time evolution of the dipole moment of
(LiH)$_2$ model, obtained using the CI space of Eq.~(\ref{eq:ormas})
with different values of $\delta$.
}%
\end{figure}

Figure~\ref{fig:dip_ormx} shows the evolution of the dipole computed
using this CI space. Unlike the simulations in
Sec.~\ref{sec:applications}, we had to use a step size control
to complete the simulation, with significantly larger
values ($\delta \geq 10^{-4}$) for the regularization parameter. 
The figure reveals the $\delta$ dependence of the computed dipoles;
although the result with $\delta^2 = 10^{-8}$ seems to be near the
convergence, the simulation with $\delta^2 = 10^{-9}$ ends up with the 
divergence at around $t = 1.63T$. This problem arises due to the near
singularity of the coefficient matrix ${\sf A}$ in Eq.~(\ref{eq:amat}),
with the smallest singular value of ${\sf A}$ dropping below $10^{-7}$
at the divergence point. Generally, the occurrence of the stability
problem is the sign that the chosen CI space is not appropriate for the
problem in hand. In the present case, the CI space should be revised as
${\sf P} \rightarrow {\sf P}^\prime = {\sf P} + \Delta{\sf P}$ with
\begin{eqnarray}
\Delta{\sf P} = \left\{[\phi_1]^2 [\phi_2]^0 + [\phi_1]^0 [\phi_2]^2 \right\}
[\phi_3\textrm{--}\phi_8]^2, 
\end{eqnarray}
which is equivalent to the HF+SD CI space,
making the rotation between $\phi_1$ and $\phi_2$ redundant. 

Another view of Fig.~\ref{fig:dip_ormx} is that the dipole obtained with
the present method (hopefully nearly convergent with respect to $\delta$)
agrees with the CAS(8) result much better than that of the HF+S method given in
Fig.~\ref{fig:dip_rhf+}. Rather the performance is similar to that of
the HF+SD method. The CI space of Eq.~(\ref{eq:ormas}) can be decomposed as
\begin{eqnarray}
\label{eq:ormas_dec}
{\sf P} &=& {\sf P}_\textrm{HF+S} +
\left[\phi_1\right]^1 \left[\phi_2\right]^1
\left[\phi_3\textrm{--}\phi_8\right]^2,
\end{eqnarray}
where ${\sf P}_\textrm{HF+S}$ is given by Eq.~(\ref{eq:rhf+_lih}) with
$L = 1$ and underlined orbitals included. Then, the aforementioned
performance comparison indicates the importance of {\it product} double
excitations represented by the second term of Eq.~(\ref{eq:ormas_dec}). 
This encourages the development of the time-dependent size extensive theory
such as the coupled-cluster theory \cite{Kvaal:2012}.

\bibliography{refs.bib}
\end{document}